\documentclass{CUP-JNL-DCE}

\usepackage{latexsym}
\usepackage{graphicx}
\usepackage{multicol,multirow}
\usepackage{amsmath,amssymb,amsfonts}
\usepackage{mathrsfs}
\usepackage{amsthm}
\usepackage{empheq}
\usepackage{apacite}
\usepackage{rotating}
\usepackage{appendix}
\usepackage[authoryear]{natbib}
\usepackage{ifpdf}
\usepackage[T1]{fontenc}
\usepackage{times}
\usepackage{newtxmath}
\usepackage{textcomp}
\usepackage{hyperref}
\usepackage{lipsum}
\usepackage{ragged2e}
\usepackage{subcaption}
\usepackage{xcolor,colortbl}

\articletype{RESEARCH ARTICLE}
\jname{Preprint submitted to Data-Centric Engineering}
\jyear{2023}

\begin{document}
	
	\begin{Frontmatter}
		
		\title{A switching Gaussian process latent force model for the identification of mechanical systems with a discontinuous nonlinearity}
		
		\author*[1]{Luca Marino}\orcid{0000-0002-9178-1695}\email{l.marino-1@tudelft.nl}
		\author[1]{Alice Cicirello}\orcid{0000-0002-6556-2149 }\email{a.cicirello@tudelft.nl}

		\address[1]{\orgdiv{Faculty of Civil Engineering and Geosciences}, \orgname{Delft University of Technology}, \orgaddress{\street{Stevinweg 1}, \postcode{2628 CN Delft}, \country{NL}}}

		
		\abstract{An approach for the identification of discontinuous and nonsmooth nonlinear forces, as those generated by frictional contacts, in mechanical systems that can be approximated by a single-degree-of-freedom model is presented. To handle the sharp variations and multiple motion regimes introduced by these nonlinearities in the dynamic response, the partially-known physics-based model and noisy measurements of the system's response to a known input force are combined within a switching Gaussian process latent force model (GPLFM). In this grey-box framework, multiple Gaussian processes are used to model the unknown nonlinear force across different motion regimes and a resetting model enables the generation of discontinuities. The states of the system, nonlinear force and regime transitions are inferred by using filtering and smoothing techniques for switching linear dynamical systems. The proposed switching GPLFM is applied to a simulated dry friction oscillator and an experimental setup consisting in a single-storey frame with a brass-to-steel contact. Excellent results are obtained in terms of the identified nonlinear and discontinuous friction force for varying: (i) normal load amplitudes in the contact; (ii) measurement noise levels and (iii) number of samples in the datasets. Moreover, the identified states, friction force and sequence of motion regimes are used for evaluating: (1) uncertain system parameters; (2) the friction force-velocity relationship and (3) the static friction force. The correct identification of the discontinuous nonlinear force and the quantification of any remaining uncertainty in its prediction enable the implementation an accurate forward model able to predict the system's response to different input forces.}
		
		\begin{policy}[Impact Statement]
			Identifying the nonlinear forces generated by frictional joints is crucial for understanding and predicting the nonlinear behaviour of engineering structures. However, most identification approaches cannot deal with the sharp variations and multiple motion regimes introduced by these nonlinearities in the system's response. The method presented in this paper combines a partially-known physics-based model of the system with noisy measurements of its response in a switching Gaussian process (GP) latent force model, where multiple GPs are used to model the nonlinear force across different motion regimes and a resetting model to generate discontinuities. Regime transitions and discontinuities are inferred in a Bayesian manner, along with the nonlinear force, and can be used to implement forward models able to make reliable predictions.
		\end{policy}
		
	\end{Frontmatter}

	\section{Introduction}
	\label{sec1}
	
	This paper focuses on one of the key challenges in structural dynamics: the identification of discontinuous nonlinear forces arising at the structural joints of complex mechanical systems, when an incomplete physics-based model of the system and noisy measurements of its dynamic response are available. Due to tighter tolerances and the general requirement for higher performance, it is becoming more and more essential to correctly account for the nonlinearity introduced by frictional joints  in structural design and analysis. The presence of friction can produce  harsh or nonsmooth phenomena such as stick-slip motions and vibro-impact, which cannot be accounted for with equivalent linear techniques even in low-amplitude vibration settings \citep{Butlin2015}.
	In particular, an approach is proposed for the identification of discontinuous and nonsmooth nonlinearities, as those introduced by dry friction in mechanical systems. 
	
	
	Nonlinearity in structural dynamics poses several challenges, including the need for advanced mathematical models and solution techniques, and the lack of a universal approach to the experimental testing of nonlinear structures \citep{Wang2018}. Nonlinear system identification plays here a fundamental role in reconciling numerical predictions with experimental investigations \citep{Kerschen2006}. In fact, it enables the extraction of information about the nonlinear structural behaviour from experimental data and, as a consequence, the  prediction of the response of these systems to different inputs. Several techniques have been developed in the recent years to deal with the detection \citep{Worden2001}, localisation \citep{Wang2018}, characterisation \citep{Ondra2017} and quantification \citep{Carella2011} of nonlinearities; for exhaustive reviews on this topic see \citep{Kerschen2006, Ewins2015, Noel2017}. Nonetheless, most nonlinear system identification approaches cannot easily handle discontinuous and nonsmooth nonlinearities, as those introduced by dry friction in mechanical systems. 
	
	A very promising approach for dealing with smooth nonlinearities is the Gaussian process latent force model (GPLFM). The GPLFM was firstly introduced by \cite{Alvarez2009} for identifying the unknown input force driving a second-order dynamical system from noisy observations of its response. The latent driving force were modelled as a zero-mean temporal Gaussian process (GP) with a stationary kernel, and  the governing equation of the system (i.e, the domain knowledge) where exploited to update the GP prior with the response measurements. Although outperforming pure data-driven approaches, this formulation was computationally expensive, since GP regression scales as $\mathcal{O}(T^3)$ with respect to the number of data points $T$. To overcome this limitation, \cite{Hartikainen2012} reformulated the problem as an augmented state-space model, coupling the governing equations of the system with the state-space representation of the GP latent force, whose derivation is presented in \cite{Hartikainen2010}. In this formulation, inference could be performed sequentially by using Kalman filter \citep{Kalman1960} and Rauch-Tung-Striebel (RTS) \citep{Rauch1965}, significantly reducing the computational burden. In recent years, the use of GPLFMs in mechanical systems has been explored by several authors. In particular, the approach was applied to linear single degree-of-freedom (SDOF) \citep{Rogers2018} and multi degree-of-freedom (MDOF) mechanical systems \citep{Nayek2019, Rogers2020}, as well as to nonlinear systems with a known nonlinearity \citep{Rogers2020-2}, to perform joint input-state estimation. In \citeauthor{Rogers2018} (\citeyear{Rogers2018}; \citeyear{Rogers2020}), the GPLFM is also used to infer the uncertain physical parameters of the system. Further applications of the GPLFM to joint input-state identification can be found in recent experimental studies \citep{Petersen2022,Zou2022,Bilbao2022,Zou2022-2}. Finally, in \cite{Rogers2022}, the GPLFM is applied to the nonlinear identification of mechanical systems with a known driving force. In the case of a SDOF Duffing oscillator, the joint estimation of system parameters, latent states and nonlinear restoring force was performed by modelling the latter as a GP latent force. In this approach, the functional form of the nonlinear force was also reconstructed by fitting the inferred estimates with a polynomial curve, whose degree was obtained by using a Bayesian Information Criterion \citep{Schwarz1978}. A limitation of GPLFMs is that a single GP latent force is generally not able to model sharp variations in the time series, as those generated by discontinuous nonlinearities or switching driving forces, or to handle the response of systems operating under different motion regimes. Different approaches have been developed for identifying systems driven by a sequence of latent forces. For example, \citeauthor{Alvarez2010} inferred the unknown switching time instants along with the GP hyperparameters. However, this approach required a prior knowledge of the number of switching points. \cite{Hartikainen2012} overcame this limitation by formulating the GPLFM as a switching linear dynamical system, where the latent driving force model transitions are governed by a discrete-time Markov model. Nonetheless, the use of switching latent force models for the identification of nonlinear systems is currently unexplored.
	
	In this paper, a switching (GPLFM) is proposed to identify a discontinuous nonlinear force and the latent states of a  single degree-of-freedom (SDOF) mechanical system excited by a known driving force. This method is based on the introduction of the switching GPLFM framework \citep{Hartikainen2012}
	in the latent nonlinear restoring force model \citep{Rogers2022}, which enables: (i) the use of different GPs to model the nonlinear forces acting under different motion regimes (e.g., during sliding and sticking responses of a dry friction oscillator); (ii) the use of resetting models for generating discontinuities in the latent nonlinear force.
	Moreover, an approach is developed for estimating the uncertain physical parameters of the system, i.e., mass, viscous damping and stiffness, from the identified latent states and discontinuous nonlinear force. Finally, a procedure is proposed for characterising the functional dependency of the nonlinear friction force on the sliding velocity and determining the static friction force. This enables the evaluation of the friction force-velocity relationship and the value of the static friction force, which are robust features for the characterization of the friction law \citep{Cabboi2022}. 
	
	
	The paper is organised as follows. The mathematical formulation of the switching GPLFM, along with its implementation, is introduced in section \ref{sec2}. The proposed methodology is applied to the numerical case-study of a SDOF dry friction oscillator under random phase multisine excitation in section \ref{sec3}. An experimental case-study is then presented in section \ref{sec4}, where nonlinear system identification is performed on the stick-slip response of a single-storey frame subject to harmonic base excitation. Finally, the results are further discussed in the concluding remarks, presented in section \ref{sec5}.
	
	\section{Mathematical formulation of the switching Gaussian process latent force model}
	\label{sec2}
\begin{figure*}
		\centering
		\begin{subfigure}{.6\textwidth}
			\centering
			\includegraphics[width=\textwidth]{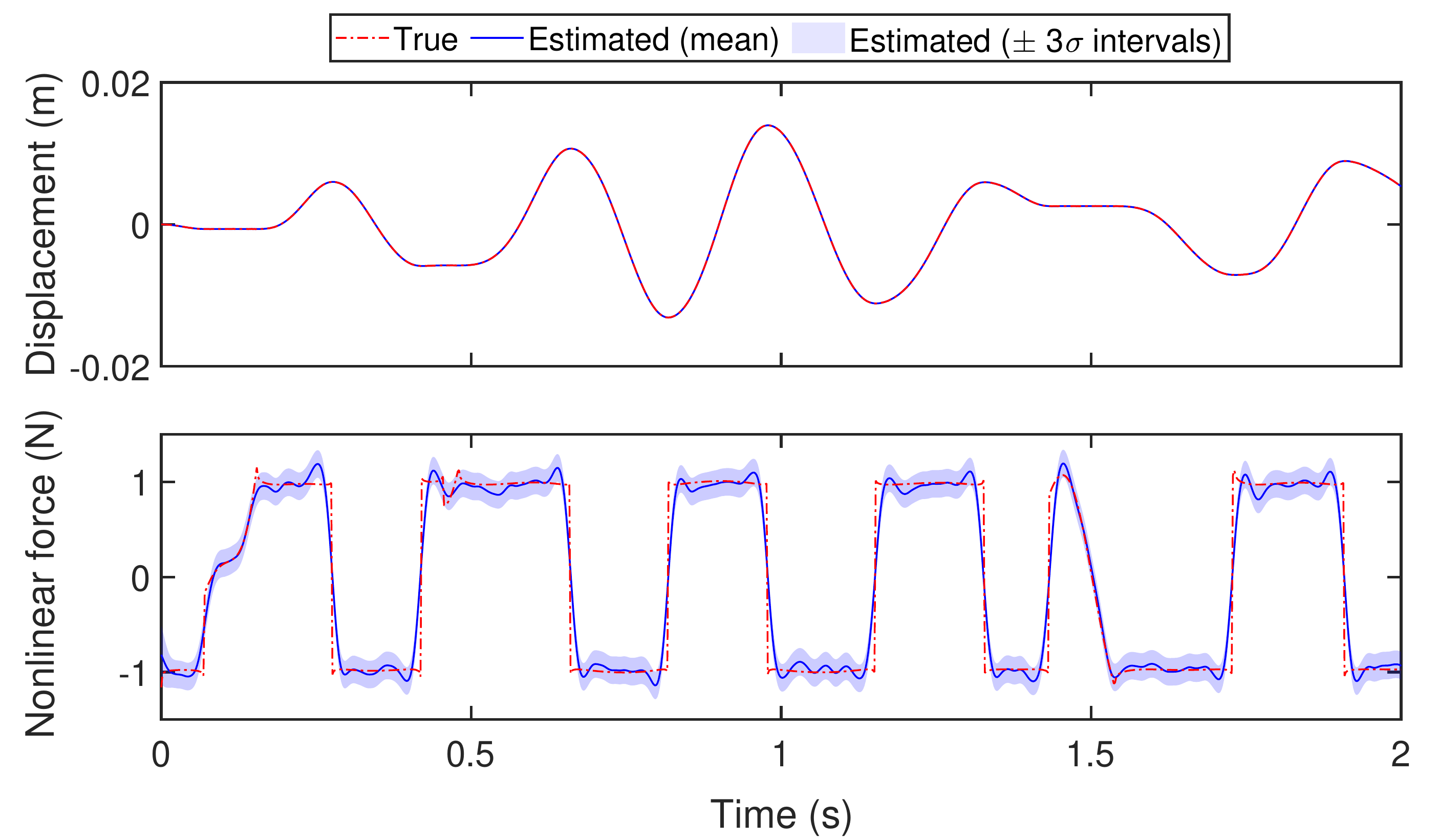}
			\caption{}
		\end{subfigure}
		\hspace{2mm}
		\begin{subfigure}{.36\textwidth}
			\centering
			\vspace{3.65mm}
			\includegraphics[width=\textwidth]{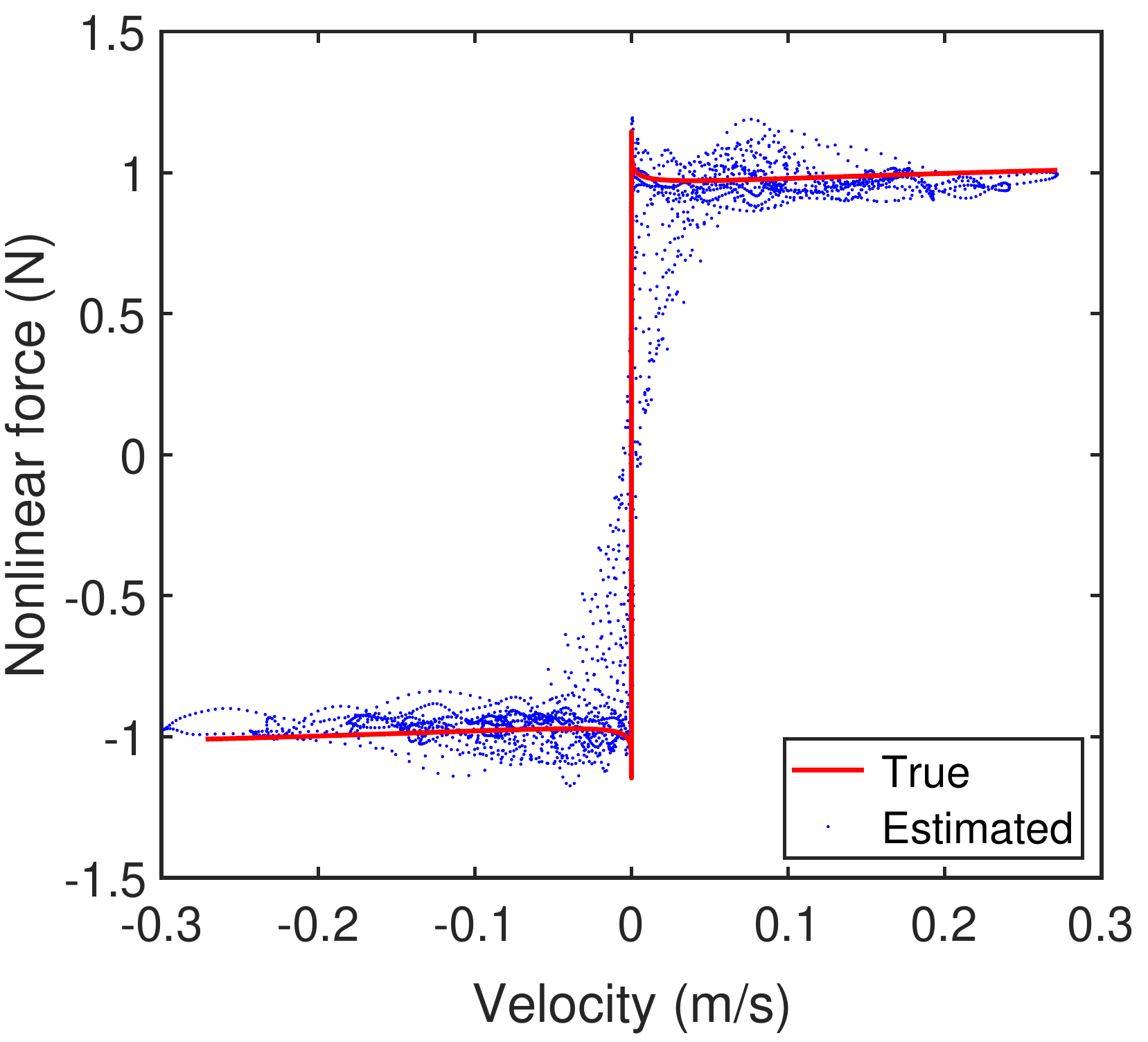}
			\caption{}
		\end{subfigure}
		\caption{Example of nonlinear restoring force identification in a dry friction oscillator (with model parameters as specified in section \ref{sec3}), obtained for $f_s=500$ Hz and SNR $= 80$ dB: displacement and nonlinear friction force time evolutions (\textbf{a}) and friction force vs velocity (\textbf{b})}
		\label{fig:1}
	\end{figure*}

	Let us consider a nonlinear SDOF system of mass $m$, viscous damping coefficient $c$ and stiffness $k$, excited by the dynamic load $u(t)$. The governing equation of this system can be expressed as:
	\begin{equation}
		m\ddot{z}+c\dot{z}+kz+f(z,\dot{z}) = u(t)
		\label{eq:14}
	\end{equation}
	where $f(z,\dot{z})$ is a generic unknown nonlinear function of the displacement and velocity of the mass. The identification of this nonlinear term, including its most plausible function form and parameters, is essential for the implementation an accurate forward model of the system able to predict its response to any input forces. If a set of noisy measurements of the system's response to a known driving force $u(t)$ is available, the nonlinear force can be identified by applying the GPLFM in the formulation proposed by \cite{Rogers2022}. In this case, the latent force model is formulated by modelling the nonlinear function as a zero-mean GP in time with a stationary kernel $\kappa$:
	\begin{equation}
		f(z,\dot{z})\sim\mathcal{GP}\left(0,\kappa(t,t')\right)
		\label{eq:15}
	\end{equation}
	The performances of this nonlinear system identification method are strictly dependent on the ability of GPs of reconstructing the nonlinear force. A particularly challenging problem for the GPLFM is the identification of nonlinear forces whose time evolution is characterised by the presence of discontinuities or sharp variations. As an example, figure \ref{fig:1} shows the performance of the GPLFM in identifying the nonlinear force in a dry friction oscillator subject to a known random phase multisine excitation, for which noisy measurements of the mass displacement are available. In this system, which is thoroughly described in section \ref{sec3}, friction always opposes the sliding motion between the parts in contact, leading to the presence of a discontinuity in the nonlinear force when the sliding velocity sign changes. In figure \ref{fig:1}a, it can be observed that the GP tends to smooth sharp variations in the identified nonlinear force. As a result, information regarding the presence of the discontinuity will be lost during the identification process. This can also be observed from the nonlinear friction force-velocity point estimates reported in figure \ref{fig:1}b. Moreover, in figure \ref{fig:1}a, it can be observed that the friction force displays significantly different patterns during the alternation of sticking and sliding phases characterising the stick-slip response of the system. If a single latent force is used to model the nonlinear friction force, this further affects the performance of the GPLFM.
	
	\subsection{Probabilistic model}
	\label{sec2.1}
	A switching GPFLM is here introduced to enable the identification of complex nonlinearities characterised by discontinuities and/or generating different motion regimes in the dynamic response. This latent force model can be formulated as a SLDS in the following form:
	\begin{subequations}
		\begin{empheq}[left = \empheqlbrace]{align}
			& p(\mathbf{x}_t|\mathbf{x}_{t-1},u_{t-1},s_t) = \mathcal{N}(\mathbf{x}_t|A(s_t)\mathbf{x}_{t-1}+B(s_t)u_{t-1},Q(s_t))\\
			& p(\mathbf{y}_t|\mathbf{x}_{t},u_t,s_t) = \mathcal{N}(\mathbf{y}_t|C(s_t)\mathbf{x}_{t}+D(s_t)u_{t},R(s_t))
		\end{empheq}
		\label{eq:19}
	\end{subequations}\\
	where $s_t$ is a switch variable denoting the active latent force model at the time instant $t$. For each model, the above equations correspond to the transition and observation models of the augmented linear Gaussian state-space model (LGSSM) obtained by coupling the state-space representations of the system and the latent force, as described in detail in section \ref{sec2.2}. 
	
	This formulation is achieved in the same spirit of that proposed by \cite{Hartikainen2012} to identify dynamical systems where different latent forces can act as an input in different, and possibly overlapping, time intervals. The proposed switching latent force model mostly differ from \citeauthor{Hartikainen2012}'s model for the presence of a known input term, which enables the use of the GP latent forces for modelling the unknown nonlinear forces acting in the system. In the switching GPLFM framework, different dynamical models can be introduced to describe the evolution of the latent states and nonlinear force at different time steps, allowing the use of different governing equations to deal with different motion regimes.
	In the example of a dry frictional oscillator, the proposed approach allows the definition of different equations to describe the behaviour of the system during sliding and sticking phases. However, to enable the presence of discontinuities in the latent nonlinear force, either at the transition between different regimes or between two sliding phases with velocities of opposite sign, it is also necessary to include a resetting model in the SLDS formulation. This model, as suggested by \cite{Hartikainen2012}, resets the latent force components to a zero-mean Gaussian prior with a suitably chosen covariance, while leaving the output states of the system unaltered. 
	
	\begin{figure*}
		\centering
		\includegraphics[width=\textwidth]{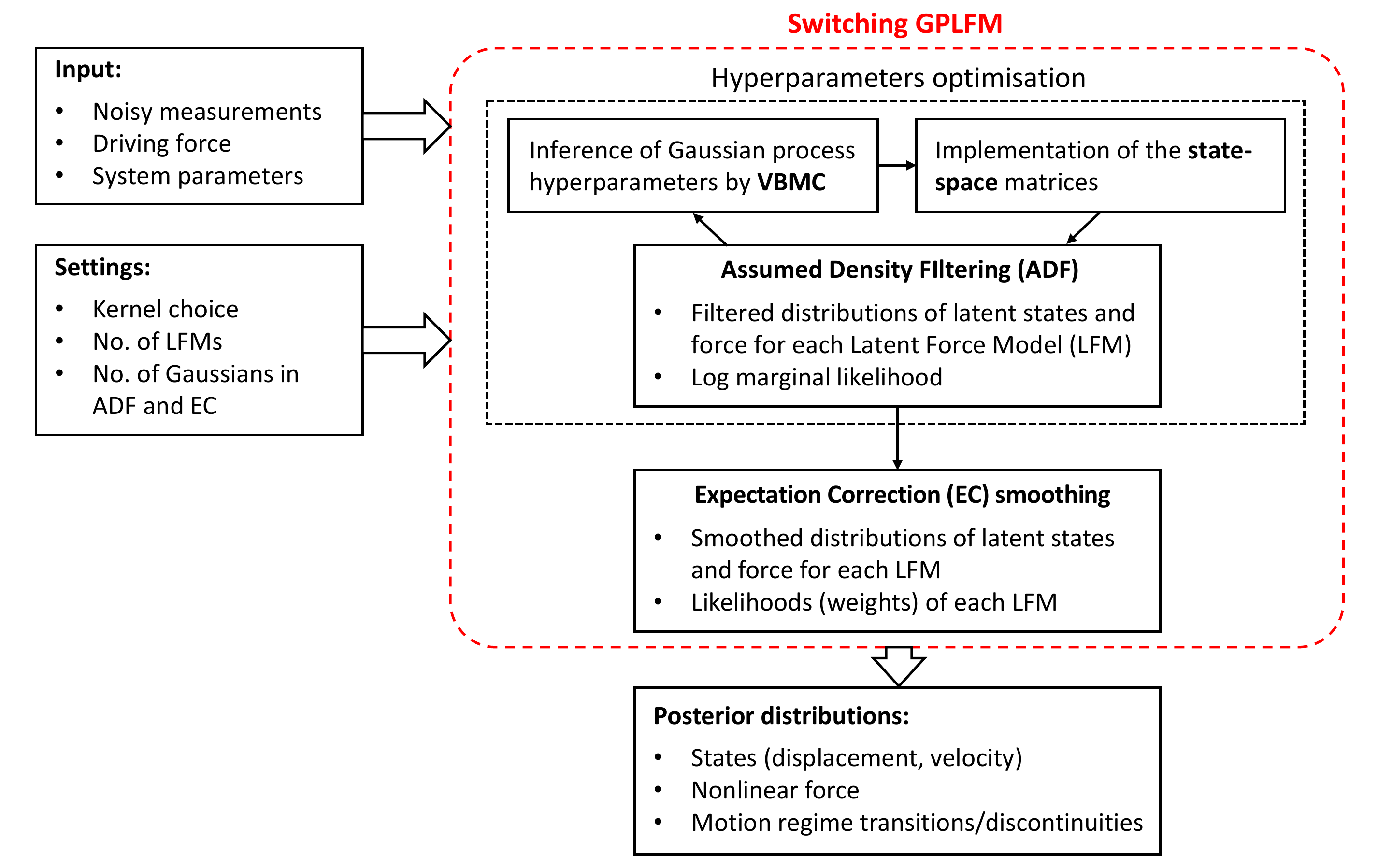}
		\caption{Schematic representation of the switching GPLFM}
		\label{scheme:1}
	\end{figure*}
	
	In the switching GPLFM formulation, the SLDS from equation (\ref{eq:19}) is generally coupled with a discrete-time Markov model of form $p(s_t|s_{t-1})$, so that probabilistic inference can also be performed on the transitions between the different latent force models. If $S$ different models are considered, with the $S$-th model operating as a resetting model, this Markov model can be written as:
	\begin{equation}
		p(s_t|s_{t-1}) =
		\begin{cases}
			\rho, & \mathrm{if}\; s_t = s_{t-1} \neq S\\
			1-\rho, & \mathrm{if}\; s_t = S \;\mathrm{and}\; s_{t-1}\neq S\smallskip\\
			\dfrac{1}{S-1}, & \mathrm{if}\; s_t \neq S \;\mathrm{and}\;  s_{t-1} = S\smallskip\\ 
			0, & \mathrm{otherwise} 
		\end{cases}
		\label{eq:20}
	\end{equation}
	with $0<\rho<1$. Equation (\ref{eq:20}) can be interpreted as follows. If a certain (non-resetting) model is active at the time $t-1$, there is a probability equal to $\rho$ that such a model will remain active at the subsequent time instant $t$; otherwise, a transition to the resetting model will occur. Furthermore, if the resetting model is active at the instant $t-1$, in the following time step there will be a transition to any of the other models, with equal probabilities. The value of $\rho$ must be chosen by the user, usually between 0.9 and 1 (see, e.g., \cite{Barber2006,Hartikainen2012}).
	
	In summary, the switching GPLFM can take into account three different sources of uncertainty in the nonlinear system identification process:
	\begin{itemize}
		\justifying
		\item \textbf{Nonlinear force model uncertainty.} The lack of knowledge of the functional form of the nonlinear term in equation (\ref{eq:14}) is addressed by modelling, for each motion regime, the unknown nonlinear force as a zero-mean temporal GP with a stationary kernel, which is then updated in a Bayesian manner by using the available response measurements and the relationship between nonlinear force and response determined by the physics-based model.
		\item \textbf{Measurement uncertainty.} As in standard GPLFMs, the measurement noise is here assumed to follow a Gaussian distribution of covariance $R$; different covariance matrices can be considered across the different latent force models. While here a zero-mean Gaussian distribution is assumed, the presence of a measurement bias could also be accounted for in the procedure; further details can be found in \cite{Barber2006}. In the proposed approach, the measurement noise covariance is inferred along with GP hyperparameters (see section \ref{sec2.4}).
		\item \textbf{Regime transitions uncertainty.} One of the main advantages of this approach is its ability to infer the unknown switches between the different latent force models, and therefore motion regime transitions and discontinuities. The prior knowledge about regime transitions is modelled by the Markov model introduced in equation (\ref{eq:20}) and updated by using the response measurements, along with the latent states and forces.
	\end{itemize}
	
	The implementation of the proposed approach is schematically shown in figure \ref{scheme:1}. In order to build the SLDS from equation (\ref{eq:19}), it is firstly necessary to derive the state-space formulations of the system and GP latent forces, according to the procedure introduced in subsection \ref{sec2.2}. Inference can finally be performed on the so-defined probabilistic model to retrieve the optimal GP hyperparameters and the posterior distributions for the latent states and nonlinear force, along with the sequence of motion regimes and discontinuities occurring in the time evolution. This is explained in subsection \ref{sec2.3}.
	
	\subsection{State-space formulation}
	\label{sec2.2}
	Let us assume that, in equation (\ref{eq:14}), the nonlinear term is expressed by a set of $S-1$ latent forces. According to the switching GPLFM formulation, one or more of these forces can be active at each time step. Each latent force can be formulated as a zero-mean temporal GP with a stationary kernel, as indicated in equation (\ref{eq:15}). As demonstrated by \cite{Hartikainen2010}, these GPs can be expressed as LGSSMs. Referring for simplicity to a single latent force $f$, this state-space representation can be expressed as:
	\begin{equation}
		\dot{\mathbf{f}}(t) = A_{c,f}\mathbf{f}(t) +L_{c,f}w(t) 
		\label{eq:2}
	\end{equation}
	where the vector $\mathbf{f} = \begin{bmatrix} f, \dot{f}, \,\ldots\,, f^{(\beta-1)} \end{bmatrix}$ is constructed by taking the latent force and its derivatives, and:
	\begin{equation}
		A_{c,f}= \begin{bmatrix} 0 &\quad 1 &  & \\ &\quad \ddots & \quad\ddots & \\ &\quad & 0 & 1 
			\\ -a_0 &\quad \ldots & -a_{\beta-2} & -a_{\beta-1} \end{bmatrix}, \qquad \qquad
		L_{c,f} = \begin{bmatrix}  & 0 & \\ &\vdots \\& 0 \\ &1 \end{bmatrix}
		\label{eq:3}
	\end{equation}
	Following the procedure described in \cite{Hartikainen2010}, the coefficients $a_0,\ldots,a_{\beta-1}$, the spectral density $q$ of the white noise process $w(t)$ and the dimensionality $\beta$ of the vector $\mathbf{f}$ can be assigned so that the latent force $f$ corresponds to the prior of a GP with the desired covariance function. The Matérn covariance functions are particularly suitable for this implementation, since they have an analytical state-space representation. The general form of this class of functions can be written as:
	\begin{equation}
		\kappa(t,t')=\sigma_f^2\:\dfrac{2^{1-\nu}}{\Gamma(\nu)}\left(\dfrac{\sqrt{2\nu}}{l}|t-t'|\right)^\nu\mathcal{K}_\nu\left(\dfrac{\sqrt{2\nu}}{l}|t-t'|\right)
		\label{eq:4}
	\end{equation}
	where $\mathcal{K}_\nu$ and $\Gamma$ are the modified Bessel and the Gamma functions respectively. $\sigma_f$ and $l$ are two hyperparameters, the first controlling the amplitude of the function, the second the lengthscale of the time variability. The smoothness parameter $\nu$ can be selected to choose the degree of smoothness of the covariance function, ranging from the exponential ($\nu=1/2$) to the squared exponential ($\nu\rightarrow\infty)$ covariances. State-space representations have also been derived for different covariance functions, including periodic and quasi-periodic kernels \citep{Solin2014} and even some non-stationary kernels \citep{Benavoli2016}. In this paper, without loss of generality, the exponential covariance function (corresponding to Matérn 1/2) will be used. In this case, equation (\ref{eq:2}) reduces to:
	\begin{equation}
		\dot{u}(t) = -\dfrac{1}{l} u(t)+w(t)
		\label{eq:5}
	\end{equation}
	where the white noise $w$ is such that its spectral density is equal to $2\sigma_f^2/l$. 
	
	The state-space representation of the GPFLM can be obtained, as suggested by \cite{Hartikainen2012}, by coupling the state-space formulation of the zero-mean temporal GP, provided in equation (\ref{eq:2}), with that of the dynamical system, which can be derived from equation (\ref{eq:14}) as:
	\begin{equation}
		\dot{\mathbf{z}}(t) = A_{c,s}\mathbf{z}(t)+B_{c,s}\left(u(t)-f(z,\dot{z})\right) = 
		\begin{bmatrix} 0 & & 1 \\ -\tfrac{k}{m} & & -\tfrac{c}{m} \end{bmatrix}
		\begin{bmatrix} z \\ \dot{z} \end{bmatrix} +
		\begin{bmatrix} 0 \\ \tfrac{1}{m} \end{bmatrix}\left(u(t)-f(z,\dot{z})\right)
		\label{eq:6}
	\end{equation}
	Introducing the augmented state vector $\mathbf{x}=\begin{bmatrix} \mathbf{z}^\top & \mathbf{f}^\top \end{bmatrix}^\top$, the coupling of equations (\ref{eq:2}) and (\ref{eq:6}) results in the augmented state-space model: 
	\begin{equation}
		\dot{\mathbf{x}}(t) = A_{c}\mathbf{x}(t)+B_{c}u(t)+L_{c}w(t)
		\label{eq:7}
	\end{equation}
	where:
	\begin{equation}
		A_{c}= \begin{bmatrix} A_{c,s} & B_{c,f} \\ \mathbf{0} & A_{c,f} \end{bmatrix}, \qquad \qquad
		B_{c}= \begin{bmatrix} B_{c,s} \\ \mathbf{0}  \end{bmatrix}, \qquad \qquad
		L_{c} = \begin{bmatrix}  \mathbf{0}  \\  L_{c,f}  \end{bmatrix}
		\label{eq:8}
	\end{equation}
	In particular, the matrix $B_{c,f}$ is obtained by adding $\beta-1$ columns of zeros to the matrix $-B_{c,s}$. The above stochastic differential equation (SDE) can finally be converted to a discrete-time dynamic model in the form:
	\begin{equation}
		\mathbf{x}_t = A\mathbf{x}_{t-1} + Bu_{t-1} +\mathbf{w}_{t-1}, \qquad 
		\mathbf{w}_{t-1}\sim\mathcal{N}(\mathbf{0},Q)
		\label{eq:9}
	\end{equation}
	where the transition matrix is obtained as $A=\exp_m(A_c\Delta t)$, being $\Delta t$ the fixed time step, and  $B = A^{-1}(A_c-I)B_c$ . The covariance matrix of the process noise in the discrete model can be written as:
	\begin{equation}
		Q = \int_{0}^{\Delta t} \exp_m(A_c(\Delta t-\tau))L_cqL_c^\top \exp_m(A_c(\Delta t-\tau))^\top\mathrm{d}\tau
		\label{eq:10}
	\end{equation}
	where $q$ the spectral density of the process noise in the continuous model. A more practical implementation of equation (\ref{eq:10})  is:
	\begin{equation}
		Q = P_\infty-AP_\infty A^\top
		\label{eq:11}
	\end{equation}
	where $P_\infty$ is the steady-state solution to the SDE describing the time evolution of the covariance of the state vector $\mathbf{z}$ and can be retrieved by solving the Lyapunov equation \citep{Sarkka2019}:
	\begin{equation}
		A_cP_\infty+P_\infty A_c^\top +L_cqL_c^T=0
		\label{eq:12}
	\end{equation}
	Finally, the transition model from equation (\ref{eq:9}) is coupled with the observation model:
	\begin{equation}
		\mathbf{y}_t = C\mathbf{x}_t+Du_t+\mathbf{v}_t, \qquad 
		\mathbf{v}_t\sim\mathcal{N}(\mathbf{0},R)
		\label{eq:13}
	\end{equation}
	to recover a full LGSSM, In the above equation, $\mathbf{y}$ is the observation vector, $C$ and $D$ the observation matrices and $R=\sigma_n^2$ the variance of the measurement noise. The choice of observation matrices depends on the typology of sensor used to perform the measurements: if displacements are measured the matrices will be $C=\begin{bmatrix} 1 & 0 & 0 & \dots & 0 \end{bmatrix}$ and $D=0$, while for accelerations that would be $C=\begin{bmatrix} -k/m & -c/m & -1/m & 0 & \dots & 0 \end{bmatrix}$ and $D = 1/m$.
	
	The LGSSM in equations (\ref{eq:9}) and (\ref{eq:13}) has been formulated, for simplicity, by referring to the standard GPLFM case, which is retrieved from the more general formulation from equation (\ref{eq:19}) for $S=1$. The procedure for deriving the SSM of the GPLFM is substantially the same when more latent force models are considered. In fact, all the matrices of the LGSSM can be determined as described above for each latent force. In general, there is no requirement for the latent forces to share the same type of kernel function, meaning that these matrices can have different dimensionality across different models. Moreover, not only the latent force, but also the governing equations of the system can differ for different values of $s_t$; this will be shown in more detail in subsection \ref{sec3.1}. Finally, the implementation of the resetting model, necessary to introduce discontinuities in the latent force, is achieved by formulating the transition matrix as $A(S)=\mathrm{blkdiag}\left(\exp_m(A_{c,s}\Delta t),\mathbf{0}\right)$ and the covariance matrix of the process noise as $Q(S)=\mathrm{blkdiag}(P_\infty-A_sP_\infty A_s^\top,P_0)$, being $P_0$ the prior covariance selected for the latent force.
	
	
	\subsection{Inference of latent states and nonlinear force}
	\label{sec2.3}
	Exact inference in the SLDS formulated in equations (\ref{eq:19}) and (\ref{eq:20}) is not computationally tractable since its complexity scales exponentially with time \citep{Bar-Shalom2001,Lerner2002}. Therefore, the filtered and smoothed posterior distributions can only be inferred in an approximated form.
	
	In this paper, approximated inference will be performed by implementing the approach proposed by \cite{Barber2006}. In this approach, the filtered and the smoothed distributions of the state of each model $s-t$, are approximated by Gaussian mixtures of $I$ and $J$ components, respectively. The number of mixture components can be chosen by the user according to the computational budget and will generally be significantly smaller than $S^t$, which is the number of Gaussian components of the exact posterior distribution at the time instant $t$.
	
	In \citeauthor{Barber2006}'s procedure, the filtered distributions are computed via assumed density filtering (ADF) \citep{Alspach1972}. The result of ADF at the time step $t$ is the Gaussian mixture:
	\begin{equation}
		p(\mathbf{x}_t|s_t,\mathbf{y}_{1:t},u_{1:t}) \cong \sum_{i=1}^I \mathcal{W}_t(i,s)\cdot\mathcal{N}(\mathbf{x}_t|\boldsymbol{\mu}_{t|t}(i,s),P_{t|t}(i,s))
		\label{eq:21}
	\end{equation}
	for the state of the model $s_t$ and an approximated expression for the model probability $p(s_t|\mathbf{y}_{1:t},u_{1:t})$. Each step of ADF requires running $IS^2$ Kalman filters.
	In a similar fashion, the smoothed distributions are obtained by implementing an expectation correction (EC) algorithm, resulting in the Gaussian mixture:
	\begin{equation}
		p(\mathbf{x}_t|s_t,\mathbf{y}_{1:T},u_{1:T}) \cong \sum_{j=1}^J \widetilde{\mathcal{W}}_t(j,s)\cdot\mathcal{N}(\mathbf{x}_t|\boldsymbol{\mu}_{t|T}(j,s),P_{t|T}(j,s))
		\label{eq:22}
	\end{equation}
	and an approximated expression for $p(s_t|\mathbf{y}_{1:T},u_{1:T})$. The EC algorithm performs $IJS^2$ RTS smoothers at each time step, and is therefore more computationally expensive then ADF. Nonetheless, it is worth mentioning that the evaluation of the marginal likelihood $p(\mathbf{y}_{1:T})$, which is required for estimating the optimal hyperparameters of the latent functions, is performed within the ADF algorithm. A detailed description of ADF and EC implementation can be found in \cite{Barber2006}.
	
	\subsection{Inference of GP hyperparameters}
	\label{sec2.4}
	The performances of switching latent force models are not only affected by the choice of appropriate kernels for the GP latent forces, but also from the selection of the system parameters and GP hyperparameters. While an estimation procedure for uncertain system parameters will be introduced in subsection \ref{sec3.5}, the problem of determining the optimal values for the hyperparameters in GPLFMs is addressed in what follows.
	
	A common approach for selecting the optimal hyperparameters of the kernel function is to use maximum likelihood estimation (see, e.g., \cite{Ghahramani1996,Nayek2019}). However, the algorithms used for determining the maximum of the likelihood function are not guaranteed to find the global maximum \citep{Petersen2022} and do not give information about the posterior distribution of the parameters. For this reason, different approaches have been proposed over the years. \cite{Zou2022} and \cite{Bilbao2022} determine the optimal hyperparameters by minimising the Hellinger distance between the empirical distribution of the measurements and the modelled Gaussian prior on the observed states; nonetheless, this approach can only be used when the response distribution is well-approximated by a Gaussian. Finally, \citeauthor{Rogers2018} (\citeyear{Rogers2018, Rogers2020, Rogers2022}) recover the full posterior distribution of the parameters by using Markov Chain Monte Carlo (MCMC). In fact, while computationally more expensive, MCMC offers the advantage of a guaranteed convergence to the true posterior distribution \citep{Gelman2013}. 
	
	In this paper, hyperparameters inference will be performed by using the Variational Bayes Monte Carlo (VBMC) method developed by \citeauthor{Acerbi2018} (\citeyear{Acerbi2018, Acerbi2020}). VBMC combines variational inference and active-sampling Bayesian quadrature to perform approximate Bayesian inference, leading to a reduced computational cost compared to MCMC. The reader is referred to the papers by \citeauthor{Acerbi2018} (\citeyear{Acerbi2018, Acerbi2020}) for a detailed description of the VBMC method. VBMC requires the iterative computation of the marginal likelihood of the SLDS, which is obtained as a by-product of ADF. Therefore, as schematically shown in figure \ref{scheme:1}, each iteration requires the implementation of the LGSSM matrices by using the current values of the hyperparameters and the application of ADF. The optimal hyperparameters are finally obtained as the mean of the posterior distribution computed by VBMC and used for inferring the latent state and nonlinear force via ADF and EC.
	
	\section{Numerical case-study: a dry-friction oscillator}
	\label{sec3}
	Identifying the friction forces generated by contacts in mechanical systems is a critical challenge in engineering. In fact, information extracted by experimental data is essential for developing and updating friction models, as well as for making predictions about the dynamic behaviour of these systems. This section presents an application of the switching GPLFM to the identification of the nonlinear friction force acting in a simulated dry friction oscillator. In particular, it is shown how the proposed methodology is not only able to identify the time evolution of the friction force, but can also be used to: (i) estimate the physical parameters of the system; (ii) reconstruct the friction force-velocity relationship; (iii) estimate the static value of the friction force. The estimated system parameters and reconstructed friction model enable the prediction of the system response to an assigned input force.
	\subsection{Simulated system}
	\label{sec3.0}
\begin{figure*}
		\centering
		\hspace{3mm}
		\begin{subfigure}{.23\textwidth}
			\centering
			\vspace{5mm}
			\includegraphics[width=\textwidth]{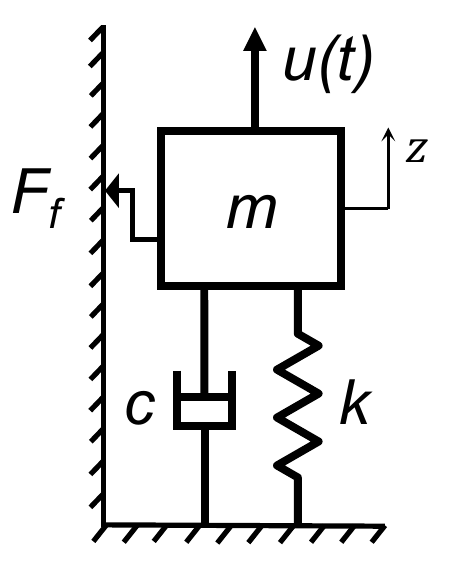}
			\vspace{-1mm}
			\caption{}
		\end{subfigure}
		\hspace{8mm}
		\begin{subfigure}{.6\textwidth}
			\centering
			\vspace{3.65mm}
			\includegraphics[width=\textwidth]{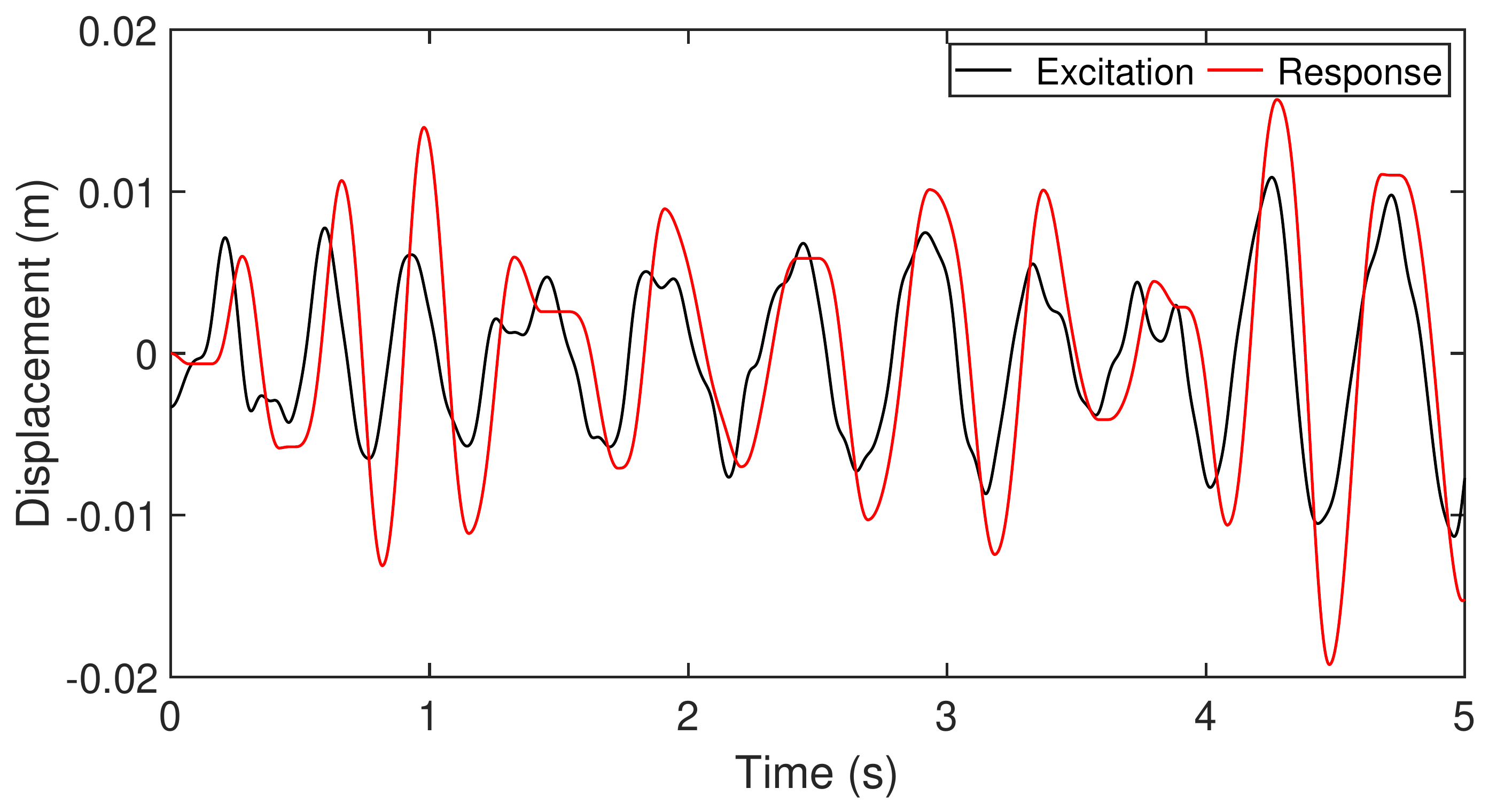}
			\caption{}
		\end{subfigure}
		\caption{Mass-spring-dashpot system with a friction contact between mass and ground-fixed wall (\textbf{a}) and its simulated response (in red) to a random phase multisine excitation (in black, divided by the stiffness) (\textbf{b})}
		\label{fig:2}
	\end{figure*}

	Let us consider the SDOF system of mass $m = 1$ kg, viscous damping coefficient $c=5$ Nsm$^{-1}$ and stiffness $k=500$ Nm$^{-1}$ shown in figure \ref{fig:2}a. A friction contact occurs between the mass and a ground-fixed wall, governed by a rate-dependent friction law, and a driving force $u(t)$ is directly applied to the mass. The governing equation of this system can be formulated as:
	\begin{equation}
		m\ddot{z}+c\dot{z}+kz+F_f(z,\dot{z}) = u(t)
		\label{eq:23}
	\end{equation}
	In particular, a steady-state version of the Dieterich-Ruina's law \citep{Dieterich1979,Rice1996}, in the formulation proposed by \cite{Cabboi2022}:
	\begin{equation}
		F_f(z,\dot{z}) =
		\begin{cases}
			\left[F_*+a\ln\left(\dfrac{|\dot{z}|+\varepsilon}{V_*}\right)+b\ln\left(c+\dfrac{V_*}{|\dot{z}|+\varepsilon}\right)\right]\mathrm{sgn}(\dot{z}), & \mathrm{if}\; \dot{z}\neq0\\
			u(t)-kz, & \mathrm{otherwise}
		\end{cases}
		\label{eq:24}
	\end{equation}
	has been used to model the friction force. The values selected for the parameters of the Dieterich-Ruina's law are reported in table \ref{tab1}. The driving force $u(t)$ is a random phase multisine whose amplitude is generated from the JONSWAP spectrum \citep{Hasselmann1973, Vazirizade2019}:
	\begin{equation}
		S(\omega) = 320\left(\dfrac{H_s}{T_p^2}\right)^2\cdot\dfrac{1}{\omega^5}\exp\left(-1.25\left(\dfrac{\omega_p}{\omega}\right)^4\right)\cdot 3.3^{\exp\,\left(\frac{1}{2\sigma_p^2}\left(1-\frac{\omega^2}{\omega_p^2}\right)\right)}
		\label{eq:25}
	\end{equation} 
	in the frequency range $\omega=(0.02:0.02:100)$ Hz, using the parameters indicated in table \ref{tab2}. The resulting input force is depicted in figure \ref{fig:2}b.
	
	It is assumed that noisy measurements of the mass displacement are available to be used as experimental observations in the GPLFM. To obtain these synthetic measurements, the dynamic response of the system has been simulated by using the event-driven variable-step Runge-Kutta (4,5) algorithm developed by \cite{Marino2022}, between the initial time instant $t=0$ and the specified final time $t=t_f$. In the simulation, the initial conditions for mass position and velocity have been set to zero. The resulting displacements, shown in figure \ref{fig:2}b, have then been resampled by using linear interpolation according to a fixed-step time vector, whose time step is obtained from the selected sampling frequency $f_s$ as $\Delta t= 1/f_s$. Finally, the resulting displacement values have been polluted with artificial white noise according to a selected signal-to-noise ratio (SNR). 
	
	In the remaining of this section, the performances of the proposed switching GPLFM approach are firstly investigated and compared to the standard latent restoring force model in subsection \ref{sec3.1}, assuming a specified set of $t_f$, $f_s$ and SNR values and known physical parameters of the system. A parameter estimation approach is then proposed in subsection \ref{sec3.2}, while the performances of the switching GPLFM for varying noise levels, observation times and sampling frequencies are investigated in subsection \ref{sec3.3}.
	
	
	\begin{table}[!h]
		\tabcolsep=0pt%
		\TBL{\caption{Selected parameters for the steady-state Dieterich-Ruina's law. \label{tab1}}}
		{\begin{fntable}
				\centering
				\begin{tabular*}{.75\textwidth}{@{\extracolsep{\fill}}cccccc@{}}\toprule%
					\TCH{$F_*\:$(N)} & \TCH{$a$} & \TCH{$b$} & \TCH{$c$} &
					\TCH{$V_*\:$(m/s)} & \TCH{$\varepsilon\:$(m/s)} \\\midrule
					\TCH{1}& 0.07 & 0.09 & 0.022 & 0.003 & 10$^{-6}$\\
					\botrule
				\end{tabular*}
		\end{fntable}}
	\end{table}
	
	\begin{table}[!h]
		\tabcolsep=0pt%
		\TBL{\caption{Selected parameters for the JONSWAP spectrum. \label{tab2}}}
		{\begin{fntable}
				\centering
				\begin{tabular*}{.75\textwidth}{@{\extracolsep{\fill}}ccccc@{}}\toprule%
					\TCH{$H_s\:$(m)} & \TCH{$T_p\:$(s)} & \TCH{$\omega_p\:$(Hz)} & \TCH{$\sigma_p\: (\omega<\omega_p)$} & \TCH{$\sigma_p\: (\omega\geq\omega_p)$} \\\midrule
					\TCH{10} & 0.5 & $2\pi/T_p$ & 0.07 & 0.09\\
					\botrule
				\end{tabular*}
		\end{fntable}}
	\end{table}
	
	\subsection{Probabilistic model for dry friction oscillators}
	\label{sec3.1}
	In order to apply the proposed switching GPLFM approach to the above numerical case-study, a probabilistic model has been implemented by considering three different latent force models ($S=3$).
	\begin{itemize} 
		\justifying
		\item A first model is introduced to describe the dynamic evolution of the system during the \textit{sliding} phases. The implementation of this model is carried out as described in detail in section \ref{sec2.2}, with the sliding friction force modelled as a zero-mean GP with an exponential kernel function.
		\item The second model aims at describing the system behaviour during the \textit{sticking} phases. In this case, the response of the system is simply characterised by constant displacement and zero velocity, while the friction force is related to displacement and forcing function as described by equation (\ref{eq:24}b). This physical knowledge of the sticking behaviour can be used to determine the prediction density as:
		\begin{equation}
			p(\mathbf{x}_t|\mathbf{y}_{1:t-1},u_{t-1},i_{t-1},s_{t-1},s_t=2) = \mathcal{N}(\mathbf{x}_t|\boldsymbol{\mu}_{t|t-1}(i_{t-1},s_{t-1}),P_{t|t-1}(i_{t-1},s_{t-1}))
			\label{eq:26}
		\end{equation}
		where:
		\begin{equation}
			\boldsymbol{\mu}_{t|t-1}(i_{t-1},s_{t-1})=\begin{bmatrix}
				\mathbb{E}(z_{t-1|t-1}(i_{t-1},s_{t-1})), 0, u_{t-1}-k\cdot\mathbb{E}(z_{t-1|t-1}(i_{t-1},s_{t-1}))
			\end{bmatrix}^\top
			\label{eq:27}
		\end{equation}
		and $P_{t|t-1}(i_{t-1},s_{t-1})=P_{t-1|t-1}(i_{t-1},s_{t-1})$. If a different kernel from Matérn 1/2 is considered, zeros can be added at the end of $\boldsymbol{\mu}_{t|t-1}(i_{t-1},s_{t-1})$ to obtain the required dimensionality. It is worth underlying that the prediction density from equation (\ref{eq:26}) can be directly filtered and smoothed in the present form, without need of defining the matrices $A(s_t)$ and $B(s_t)$ for the sticking model.
		\item The third model is the \textit{resetting} model, which is implemented as explained in subsection \ref{sec2.2}, setting the prior covariance of the latent force to $P_0=0.05$. When activated, this model resets the latent friction force to zero, with $P_0$ as covariance), while leaving displacement and velocity unaltered. Since, as described by equation (\ref{eq:20}), this model can only remain active for a single time step, the result is that the values assumed by the friction force immediately before and after the reset will not directly affect each other. This allows for the presence of a discontinuity in the friction force, either between a sticking and a sliding phases or between two sliding phases with different velocity sign.
	\end{itemize}
	Finally, the model transitions are governed by the Markov model introduced in equation (\ref{eq:20}), where $\rho$ has been set to 0.92.
	
	\begin{figure*}
		\centering
		\includegraphics[width=0.9\textwidth]{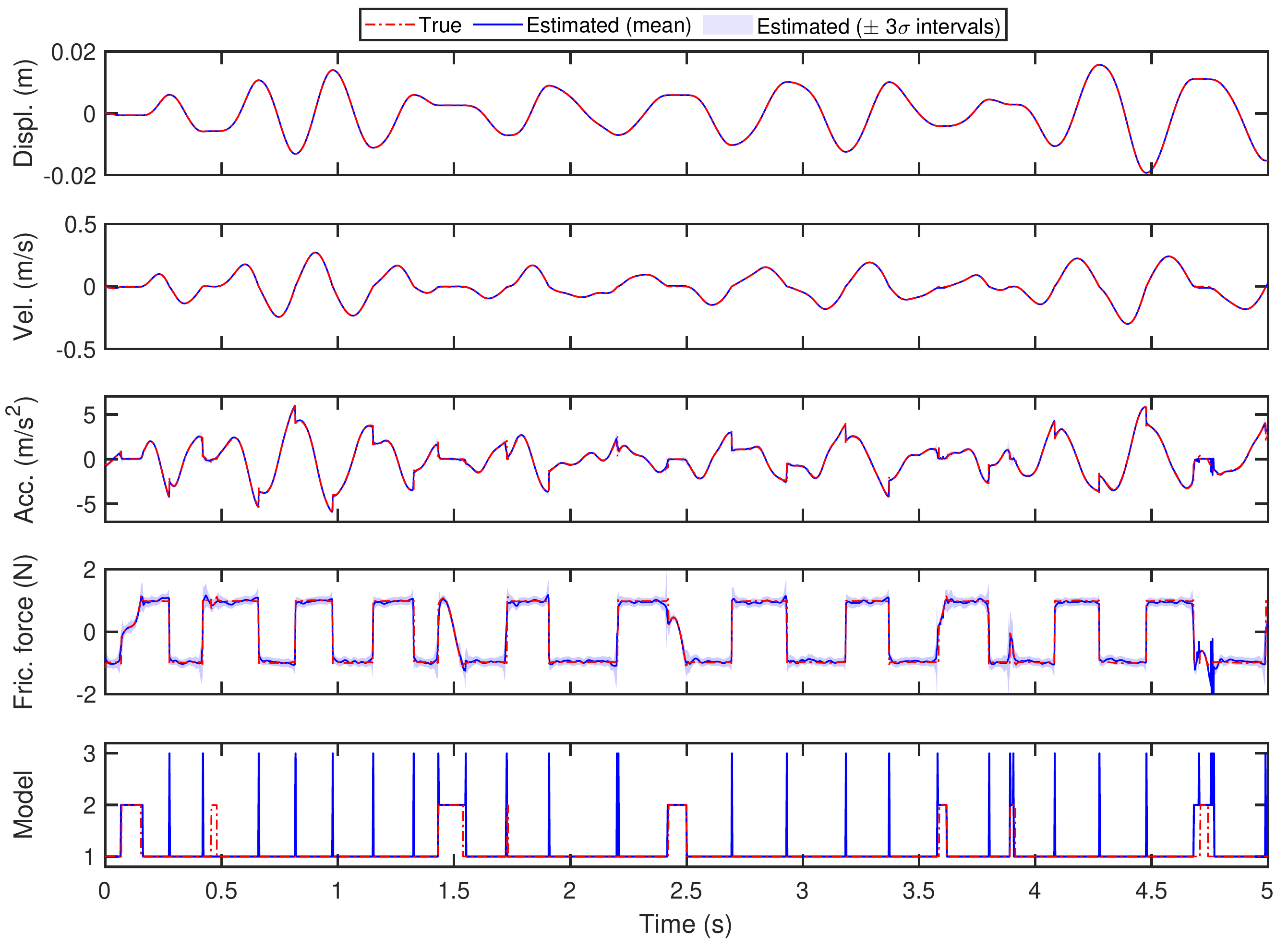}
		\caption{Latent states, acceleration, nonlinear friction force and model sequence inferred by switching GPLFM ($I=J=3$) vs ground truth. Models 1,2 and 3 stands for sliding, sticking and resetting models, respectively}
		\label{fig:3}
	\end{figure*}
	\begin{figure*}
		\centering
		\includegraphics[width=0.9\textwidth]{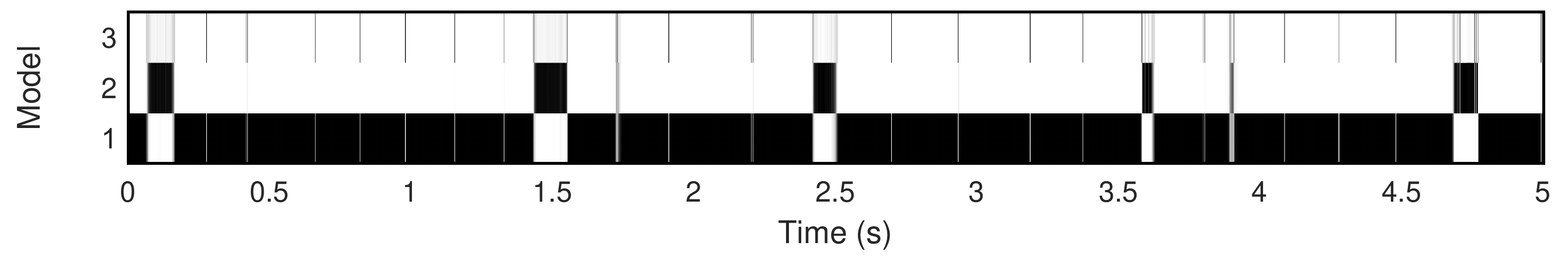}
		\caption{Model probabilities estimated by switching GPLFM ($I=J=3$). Models 1,2 and 3 stands for sliding, sticking and resetting models, respectively}
		\label{fig:4}
	\end{figure*}
	
	\subsection{Inference of latent states and nonlinear force with known system parameters}
	\label{sec3.2}
	Nonlinear force identification has firstly been performed by considering the simulated dynamic response of the system introduced above for $t_f=5\,$s and $f_s=500$ Hz. The mass displacement has been corrupted with artificial white noise by using the AWGN function of Matlab \cite{}, setting SNR $=80$. The so-produced white noise has a standard deviation of $7.8623\times10^{-3}\,$mm, corresponding to a variance $\sigma_n^2=6.1815\times10^{-5} \mathrm{mm}^2$.
	
	\begin{table}[!h]
		\tabcolsep=0pt%
		\TBL{\caption{Prior distributions for the GP hyperparameters used in the numerical case-study. \label{tab3}}}
		{\begin{fntable}
				\centering
				\begin{tabular*}{.75\textwidth}{@{\extracolsep{\fill}}ccc@{}}\toprule%
					\TCH{$\sigma_f^2\:$(N$^2$)} & \TCH{$l\:$(s)} & \TCH{$\sigma_n^2\:$(mm$^2$)} \\\midrule
					$\mathcal{N}(20,100)$ & $\mathcal{N}(20,100)$ & $\mathcal{N}(2\times10^{-5},10^{-10})$\\
					\botrule
				\end{tabular*}
		\end{fntable}}
	\end{table}
	
	In this example, the system parameters have been assumed as known a priori and set to their true values indicated at the beginning of this section. The posterior distributions of the hyperparameters $\sigma_f^2$, $l$ and $\sigma_n^2$, whose assigned priors are reported in table \ref{tab3}, have been inferred by using VBMC. As specified in subsection \ref{sec2.3}, the log likelihood function is here obtained as by-product of the ADF. Once the optimal GP hyperparameters have been estimated as the mean values of their posterior distributions, the latent states and nonlinear restoring force have been inferred by implementing ADF and EC. The number of Gaussian mixture components in the filtering and smoothing passes has here been set to $I=J=3$. The estimated displacement, velocity and friction force are shown in figure \ref{fig:3}, where it is also possible to observe their comparison with the ground truth. The agreement between the mean of the inferred distribution and the ground truth can be quantified by evaluating the normalised mean squared error (NMSE), which can be defined as:
	\begin{equation}
		\mathrm{NMSE}[g] = \dfrac{1}{T\sigma_g^2}\sum_{t=1}^T(g_t-\mathbb{E}(\hat{g}_t))^2
		\label{eq:28}
	\end{equation} 
	where $g_t$ and $\mathbb{E}(\hat{g}_t)$ are the ground truth and the estimated mean value of a certain quantity $g$ at the time step $t$, respectively, $\sigma_g^2$ is the variance of $g$ and $T$ is the number of data points. According to the above definition, a value of zero corresponds to a perfect fit between true and estimated functions, while a unitary value is obtained if $\mathbb{E}(\hat{g}_t)$ is equal to the mean value of the ground truth at every point. The very low NMSE scores reported in table \ref{tab4} for displacement and velocity confirm their excellent agreement with the ground truth. The nonlinear force identification score is NMSE$[F_f]=1.95\%$, which is generally considered a very good result. Also, the acceleration, which is determined as a linear combination of states and friction force, presents a very good agreement with the ground truth (NMSE$[\ddot{z}]=0.46\%$), despite the presence of discontinuities in its time evolution. Finally, in the bottom frame of figure \ref{fig:3}, it can be noted that most discontinuities and regime transitions are well captured by the model transitions in the switching GPLFM. Among the seven stops occurring in the observed time window, only the second has been missed; however, it can be noted that this sticking phase is particularly short and does not affect significantly the friction force value. Similarly, spikes corresponding to the resetting model activation can be observed every time that discontinuities occur between the sliding stages. It is worth remembering that the switching GPLFM frameworks allows for more latent force models to be active simultaneously and, therefore, the blue line in figure \ref{fig:3} is only referred to the predominant model at each time step. A full picture of the model probabilities is instead provided in figure \ref{fig:4}. Here, it can be seen that the resetting model is also activated in the transitions between sliding and sticking phases, although its probability does not always overcome that of the sticking model.
	
	
	\begin{table}[!h]
		\tabcolsep=0pt%
		\TBL{\caption{Latent states, acceleration and friction force identification error scores and optimal hyper- parameters for standard and switching GPLFMs with vary number of Gaussian mixture components. The ground truth for the measurement noise variance is $\sigma_n^2 = 6.182\times10^{-11}m^2$. \label{tab4}}}
		{\begin{fntable}
				\centering
				\small
				\begin{tabular*}{\textwidth}{@{\extracolsep{\fill}}lcccccccc@{}}\toprule%
					& NMSE[$z$] & NMSE[$\dot{z}$] & NMSE[$\ddot{z}$] &  \textbf{NMSE[$F_f$]} & \textbf{NMV[$F_f$]} & $\;\;\quad \hat{\sigma}_f^2\quad\;\;$ & $\;\;\quad \hat{l}\quad\;\;$ & $\hat{\sigma}_n^2$ \\
					& (\textpertenthousand) & (\%) & (\%) & \textbf{(\%)} & \textbf{(\%)} & (N$^2$) &  (s)  & ($10^{-11}$m$^2$) \\
					\midrule
					Standard &  0.0107 & 0.0024 & 0.7665 & \textbf{3.2705} & \textbf{2.8473} & 3.6567 & 0.4169 & 7.188\\
					Switching ($I=J=3$) & 0.0111 & 0.0223 & 0.4561 & \textbf{1.9507} & \textbf{0.5393} & 10.19 & 27.02 & 6.531 \\
					Switching ($I=J=5$) & 0.0114 & 0.0198 & 0.3863 & \textbf{1.6552} & \textbf{0.4722} & 4.373 & 52.77 & 7.218 \\
					Switching ($I=J=7$) & 0.0119 & 0.0200 & 0.3886 & \textbf{1.6627} & \textbf{0.5097} & 1.051 & 51.51 & 6.948 \\
					Switching ($I=J=9$) & 0.0115 & 0.0188 & 0.4754 & \textbf{2.0348} & \textbf{0.6963} & 3.860 & 73.93 & 7.079 \\
					\botrule
				\end{tabular*}
		\end{fntable}}
	\end{table}
	
	\subsection{Comparison between standard and switching GPLFMs}
	\label{sec3.3}
	In this subsection, the performance of the switching GPLFM in latent states and nonlinear force identification are investigated for varying number of Gaussian mixture components. For simplicity, the same number of Gaussians is used for ADF and EC $(I=J)$, although the EC implementation, as proposed by \cite{Barber2006}, also allows for a smaller number of Gaussians in the smoothed distribution ($J<I$). The standard GPLFM, which can simply be seen as a particular case of switching GPLFM for $S=1$, $I=1$ and $J=1$, will be here considered as a reference case. In this case, ADF and EC will automatically reduce to Kalman filtering and RTS smoothing, respectively.
	
\begin{figure*}
		\centering
		\begin{subfigure}{0.95\textwidth}
			\centering
			\includegraphics[width=\textwidth]{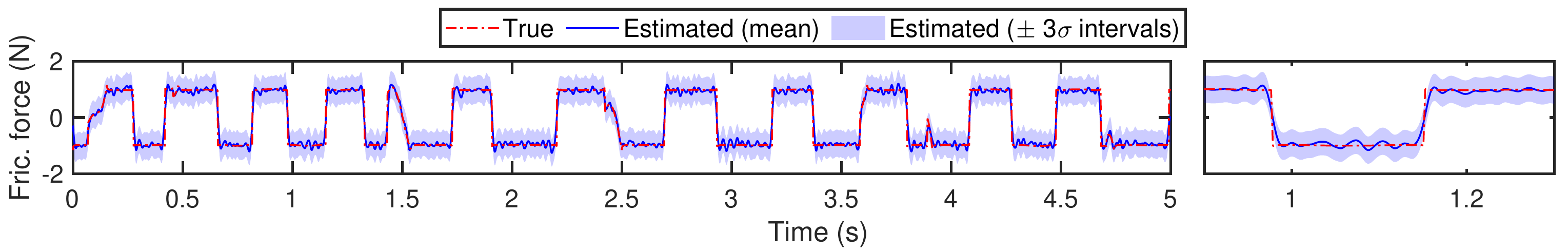}
			\caption{$S=1$, $I=1$, $J=1$}
		\end{subfigure}\\
		\begin{subfigure}{.95\textwidth}
			\centering
			\includegraphics[width=\textwidth]{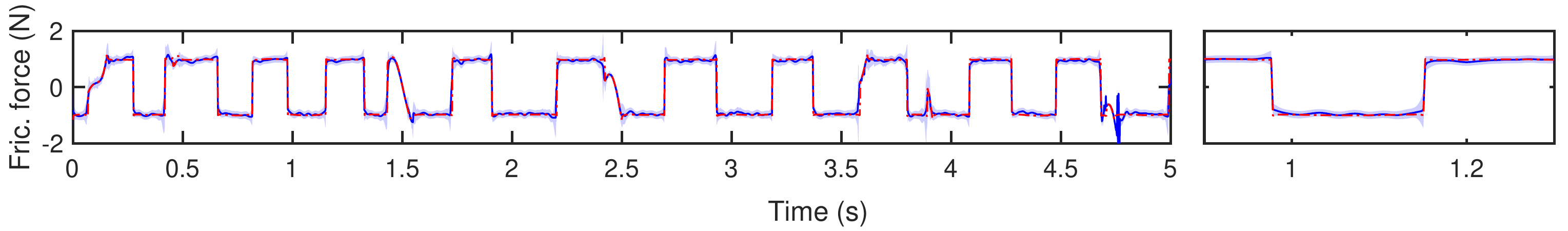}
			\caption{$S=3$, $I=3$, $J=3$}
		\end{subfigure}\\
		\begin{subfigure}{.95\textwidth}
			\centering
			\includegraphics[width=\textwidth]{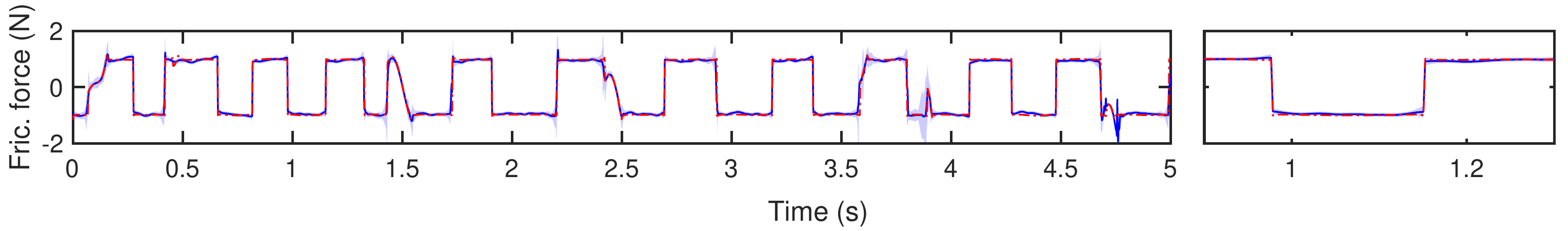}
			\caption{$S=3$, $I=5$, $J=5$}
		\end{subfigure}\\
		\begin{subfigure}{.95\textwidth}
			\centering
			\includegraphics[width=\textwidth]{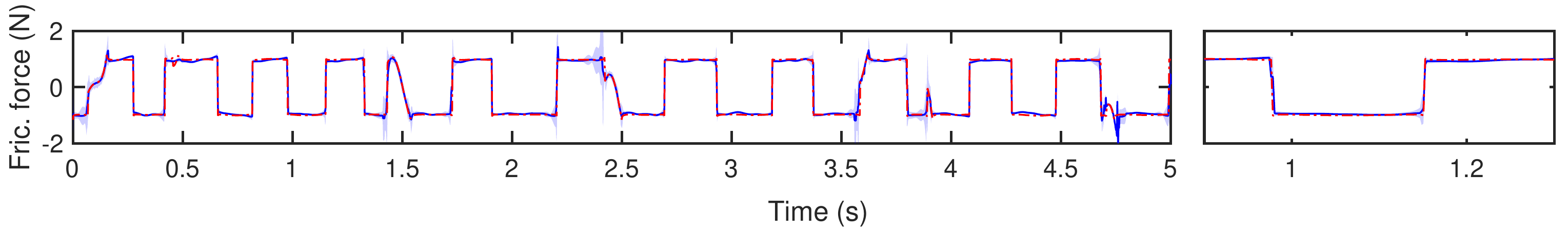}
			\caption{$S=3$, $I=7$, $J=7$}
		\end{subfigure}\\
		\begin{subfigure}{.95\textwidth}
			\centering
			\includegraphics[width=\textwidth]{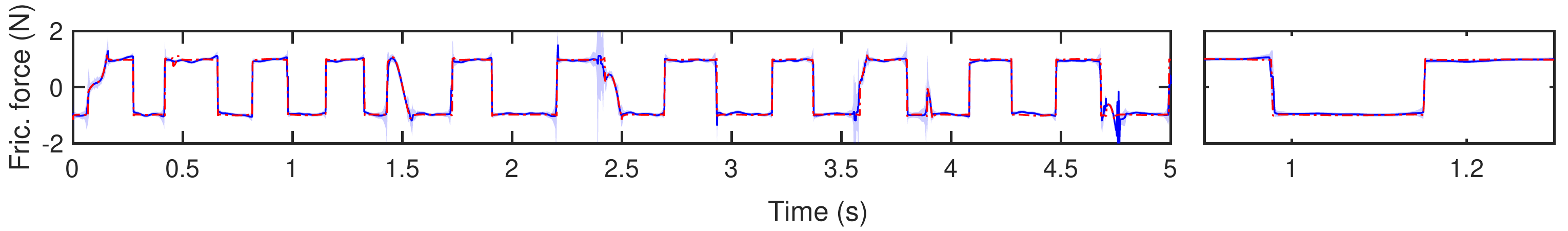}
			\caption{$S=3$, $I=9$, $J=9$}
		\end{subfigure}
		\caption{Nonlinear friction force inferred by standard and switching GPLFMs for varying number of Gaussian components in ADF and EC ($I=J$) vs ground truth}
		\label{fig:5}
	\end{figure*}

	The latent states and nonlinear force identification errors obtained by using standard and switching GPLFMs (with 3, 5, 7 and 9 Gaussian components) are reported in table \ref{tab4}, along with the optimal GP hyperparameters. Among the error scores, the normalised mean variance (NMV) index has been introduced as:
	\begin{equation}
		\mathrm{NMV}[g] = \dfrac{1}{T\sigma_g^2}\sum_{t=1}^T\mathbb{V}(\hat{g}_t)
		\label{eq:29}
	\end{equation} 
	According to this definition, a zero value is obtained if the variance of the predicted function is zero at every point, while a unitary NMV corresponds to a predictive function whose variance at every time step $t$ is equal to the overall variance of the ground truth. The identified nonlinear forces are also shown in figure \ref{fig:5}, where they are graphically compared to the true friction force in the time domain; a small section of the observed time window is enlarged on the right to illustrate how the different GPLFM approaches perform at discontinuities.
	
	Figure \ref{fig:5}a shows a good overall agreement between true and identified friction force obtained by using standard GPLFM. While discontinuities cannot be captured by a single latent force, the mean prediction follows quite closely the ground truth, resulting in an identification error of NMSE$[F_f]=3.27\%$. On the other hand, it can be observed that the confidence bounds around the mean prediction remain quite large (NMV$[F_f]=2.85\%$). The optimal hyperparameters reported in table \ref{tab4} suggest that the good agreement between the inferred and true friction force has been obtained by reducing the length-scale hyperparameter of the kernel function. In fact, small values of $l$ allow for faster variations of the latent function, increasing its capability to adapt to the sharp variations of the friction force. Therefore, this partially compensate the inability of a single GP latent force to model discontinuities and different motion regimes (i.e., sliding and sticking phases). Nonetheless, short length-scales are also associated with well-known downsides, such as the increase of model complexity and a more pronounced tendency to overfitting, particularly if noisy inputs are considered \citep{Rasmussen2006, Beckers2021}.
	
	In the remaining frames of figure \ref{fig:5}, it can be observed that the introduction of the sticking and resetting models enable the capture of most discontinuities and transitions between different motion regimes in the friction force. This leads to a significant improvement of the nonlinear force identification performance: as reported in table \ref{tab4}, the NMSE score associated with the friction force drops to $1.95\%$ when three Gaussian components are considered and even to $1.66\%$ when $I=J=5$. Nonetheless, further increases in the number of components does not appear to lead to further improvements and NMSE has been observed to increase above $I=J=7$. A possible explanation of this trend can be found in the presence of some increasingly irregular transition between motion regimes as the number of Gaussians is increased (see, for example, the transition at $t\cong 2.4\,$s). This result is also reflected by the NMV score; nonetheless, the standard deviations of the friction force are overall significantly smaller compared to the case of the standard GPLFM. Finally, looking at the optimal hyperparameters, it can be noted that the length-scales obtained by introducing the switching GPLFM are significantly larger compared to the value obtained for the standard latent restoring force model, while overall good performances are observed in the measurement noise estimation independently of the number of Gaussian components.
	
	\subsection{Nonlinear friction force characterisation}
	\label{sec3.4}
	In the previous subsections, the application of the switching GPLFM to the simulated response of the dry friction oscillator has enabled the identification of the time evolution of the nonlinear friction force, state and regime transitions. One of the advantages of GPLFMs is that no specific assumptions are required regarding the functional form of the latent nonlinear force. In this case-study, basic physical constraints have been considered in the implementation of the sticking regime, by imposing that stops are characterised by a constant displacement and zero velocity. However, no assumptions have been made about the dependence of the friction force on displacement and velocity in sliding conditions.
	Here it is shown how the identified friction force and state, along with the inferred stick-slip regime transitions, can be used to reconstruct the underlying friction law. This law, along with a correct estimation of the system parameters (dealt within section \ref{sec3.5}), enables the implementation of a robust forward model which can be used to predict the response of the dry friction oscillator to different inputs. The reconstruction of the friction model includes the determination of (i) the friction force-velocity relationship in sliding conditions and (ii) the value of the static friction force. These two steps are presented in what follows.
	
	The characterisation of the friction force-velocity relationship, or more in general of the sliding friction model, can simply be performed by fitting the latent force-latent states estimates. Several different approaches may be used for fitting; for instance, the parameters of an existing friction model can be estimated by minimising the least squared error, or a black-box approach, such as a neural network or a further GP, can be considered. For instance, in \cite{Rogers2022}, where a standard GPLFM is used for the identification of a Duffing oscillator, the nonlinear term is characterised by using a Bayesian Information Criterion to establish the most likely polynomial order of the nonlinear force-displacement relationship. Here, it has been chosen to fit the nonlinear force-velocity estimates inferred by the switching GPLFM ($I=J=5$) with the steady-state Dieterich-Ruina's law introduced in equation (\ref{eq:24}a). This choice is motivated by the flexibility of this friction model in describing, for varying parameters, both the velocity-weakening and velocity-strengthening behaviours typically observed in experimental tests \citep{Cabboi2022}; obviously, different laws or functions could be considered. The optimal values of the parameters $a$, $b$, $c$ and $F_*$ have been determined by minimising the least squared error, while $V_*$ and $\varepsilon$ have been set to the values reported in table \ref{tab1}.
	
	\begin{figure}[h!]
		\centering
		\hspace{3mm}
		\begin{subfigure}{.46\textwidth}
			\centering
			\includegraphics[width=\textwidth]{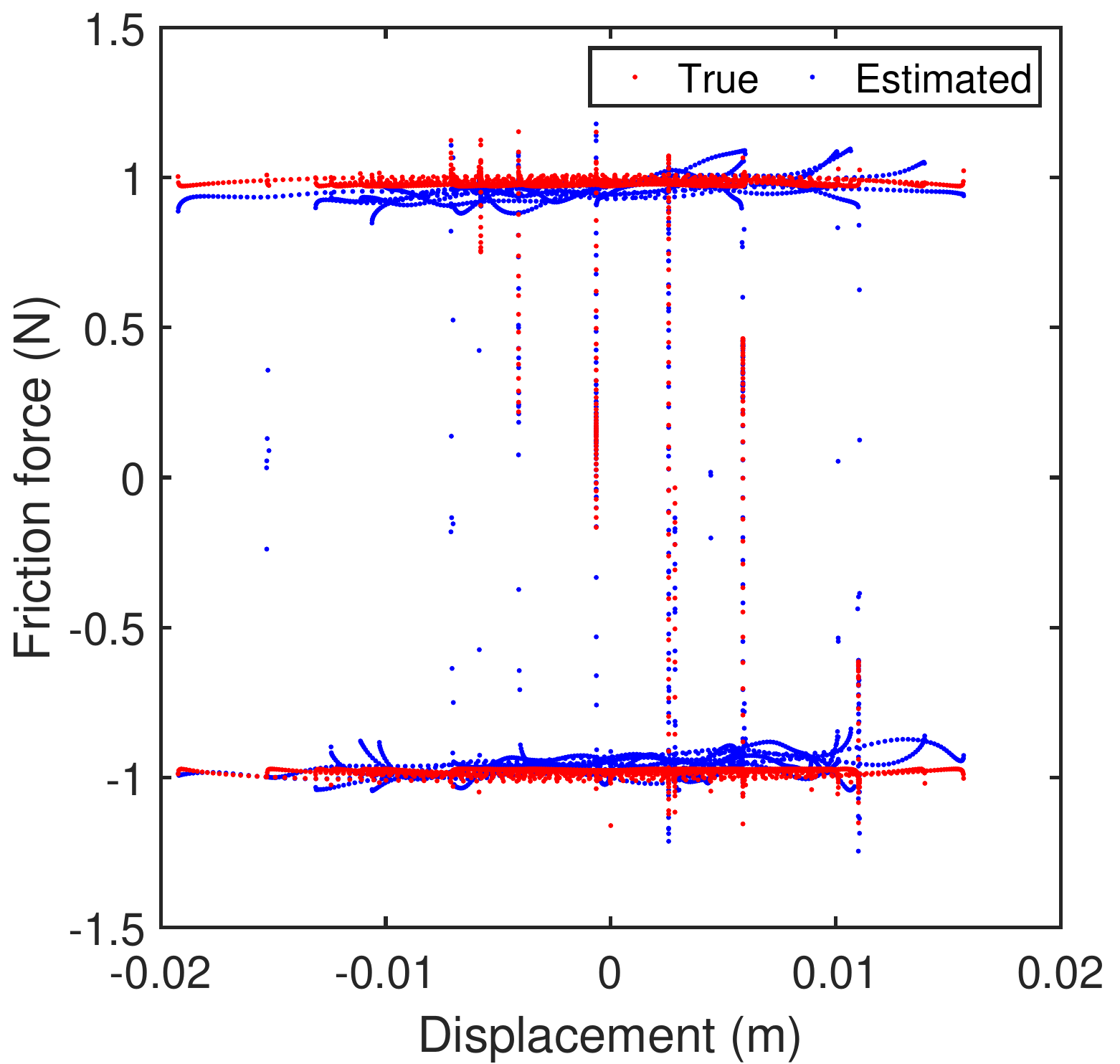}
			\caption{}
		\end{subfigure}
		\hspace{3mm}
		\begin{subfigure}{.46\textwidth}
			\centering
			\includegraphics[width=\textwidth]{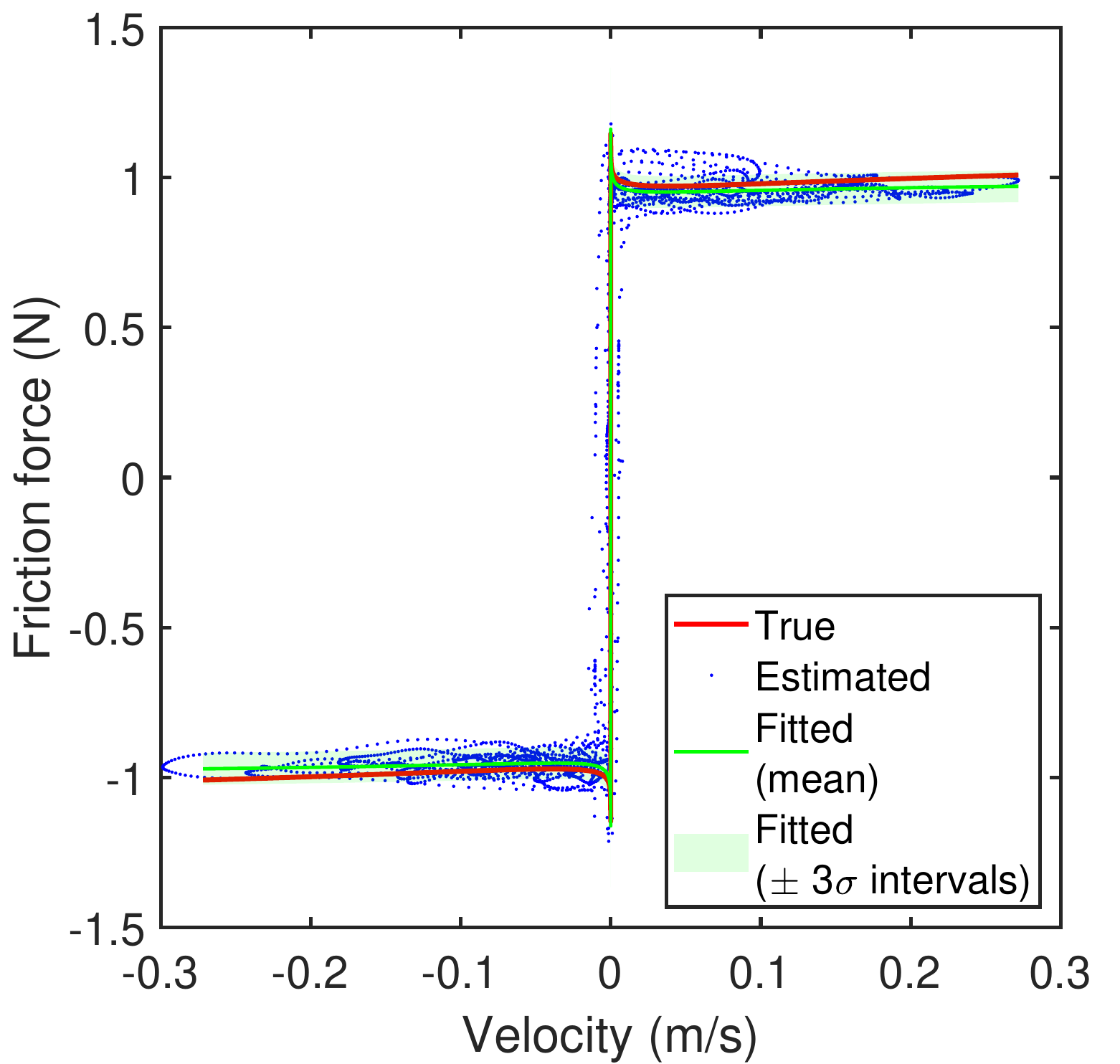}
			\caption{}
		\end{subfigure}
		\caption{Nonlinear friction force vs mass displacement (\textbf{a}) and velocity (\textbf{b}): comparison between simulated (in red) and inferred (in blue) values. The fitted friction force-velocity curve is also reported (in green) in (\textbf{b})}
		\label{fig:6}
	\end{figure}

\begin{figure}[h!]
		\centering
		\includegraphics[width=0.6\textwidth]{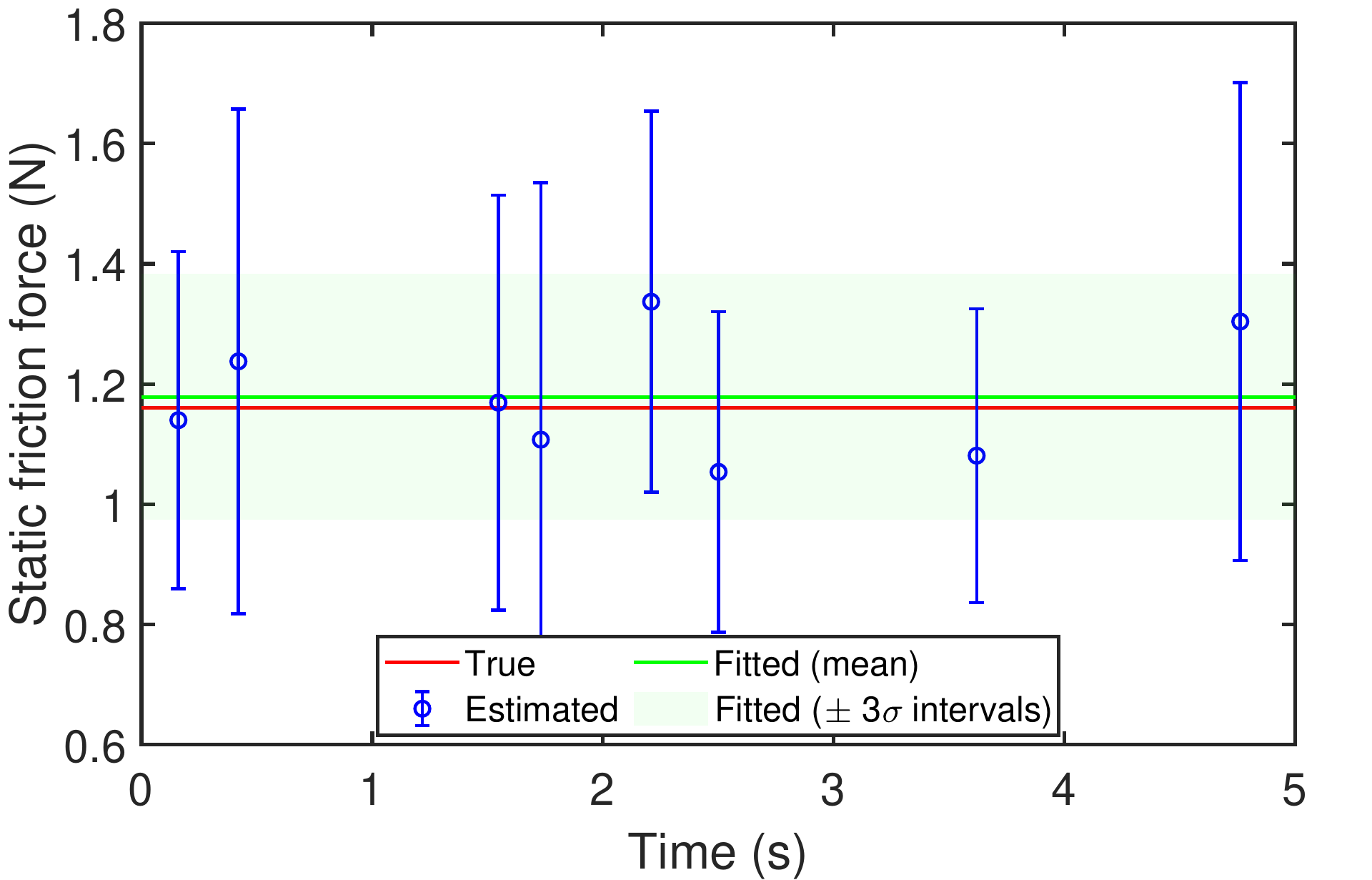}
		\caption{Estimation of the static friction force from the switching GPLFM results.}
		\label{fig:7}
	\end{figure}

	While the above fitting procedure could also be applied to estimates from the standard GPLFM, the information provided by the switching GPLFM about regime transitions also enables the estimation of the static friction force. In fact, it is well-known that the transitions from sticking to sliding regime take place when the instantaneous absolute value of the friction force value equates the static friction force (see \cite{Cabboi2022} for reference). Therefore, following the model sequence illustrated in figure \ref{fig:3}, it is possible to use the absolute values of the latent force in correspondence of these transitions as estimates of the static friction force. To finalise the procedure, the mean and the standard deviation of these estimates are evaluated, as graphically shown in figure \ref{fig:7}. The estimated static friction force $\hat{F}_s$ can be included in the fitted friction model by imposing $F_f(z,0)=\hat{F}_s$ in equation (\ref{eq:24}a). This position results in the constraint:
	\begin{equation}
		b = a + \dfrac{\hat{F}_s-F_*}{\ln(V_*)-\ln(\varepsilon)}
		\label{eq:30}
	\end{equation} 
	leaving only $a$, $c$ and $F_*$ as independent parameters of the fitting model. Figure \ref{fig:7} illustrates the comparison of the estimated static friction force, reported along with its confidence bounds, and the true value selected in the numerical simulation. It can be observed how those values, reported for $I=J=5$ are very close, with the ground truth falling well within the bounds. It is worth underlining that the accuracy of this procedure strongly depends on the observed number of stops; in a continuously sliding motion, it will not be possible to make any predictions on the static friction value.
	
	\subsection{Parameter estimation}
	\label{sec3.5}
	In the previous analyses, the mass, the viscous damping coefficient and the spring stiffness of the system have been assumed as known. However, it is often the case in experimental contexts that the values of these physical parameters are not exactly known.  In several applications of the GPLFM to mechanical systems (see, e.g., \citeauthor{Rogers2020} (\citeyear{Rogers2020,Rogers2022})), the uncertain system parameters are inferred, along with the GP hyperparameters, by using sampling approaches such as MCMC. This approach also requires to provide a prior distribution for these parameters, which is usually specified based on engineering and physical knowledge. Nonetheless, as highlighted by \cite{Rogers2022}, the presence of a latent nonlinear force can introduce a bias in the parameter estimation. For instance, the presence of a Duffing-type nonlinearity is likely to lead to incorrect linear stiffness estimates. In return, incorrect parameter estimates will also alter the identified nonlinear force.
	
	In the case of a dry friction oscillator, it has been observed that parameter inference, performed by using VBMC, does not lead to accurate estimates when either standard or switching GPLFM is applied, even if informative prior distributions are provided. Therefore, it can be concluded that, in this case, there is no real advantage in performing parameter inference rather than simply assuming initial guesses based on engineering insights. A procedure for correcting these initial guesses, based on the inferred latent force and the physical knowledge of the system, is presented in what follows.
	
	Let us denote with $m$, $c$, $k$ the true values of the system parameters, and with $\hat{m}$, $\hat{c}$, $\hat{k}$ the initial guesses, so that:
	\begin{equation}
		m = \hat{m} + \Delta m \qquad\qquad 	c = \hat{c} + \Delta c
		\qquad\qquad 	k = \hat{k} + \Delta k
		\label{eq:31}
	\end{equation} 
	By substituting equation (\ref{eq:31}) into equation (\ref{eq:23}), it can be seen how using incorrect parameter values will introduce additional linear forces in the governing equation of the system:
	\begin{equation}
		\hat{m}\ddot{z}+\hat{c}\dot{z}+\hat{k}z+\underbrace{F_f+\Delta m\cdot\ddot{z}+\Delta c\cdot\dot{z}+\Delta k\cdot z}_\text{$F_L$} = u(t)
		\label{eq:32}
	\end{equation}
	Therefore, if the latent states $z$, $\dot{z}$ and the latent force $F_L$ are inferred by considering incorrect initial guesses as system parameters, these additional forces will become part of the latent force, alongside the nonlinear friction force. The acceleration $\ddot{z}$, which is not directly inferred when applying the GPLFM, can be retrieved, from the above equation, as:
	\begin{equation}
		\ddot{z} = \dfrac{1}{\hat{m}}\left(u(t)-\hat{k}z-\hat{c}\dot{z}-F_L\right)
		\label{eq:33}
	\end{equation}
	Substituting equation (\ref{eq:33}) into the expression of the latent force $F_L$ and rearranging, it is obtained that:
	\begin{equation}
		F_L = \left(\dfrac{\hat{m}}{\hat{m}+\Delta m}\right)F_f+\left(\dfrac{\hat{m}}{\hat{m}+\Delta m}\Delta k-\dfrac{\Delta m}{\hat{m}+\Delta m}\hat{k}\right)z+\left(\dfrac{\hat{m}}{\hat{m}+\Delta m}\Delta c-\dfrac{\Delta m}{\hat{m}+\Delta m}\hat{c}\right)\dot{z}+\left(\dfrac{\Delta m}{\hat{m}+\Delta m}\right)u
		\label{eq:34}
	\end{equation}
	The above equation highlights how incorrect physical parameters introduce a linear dependency of the inferred latent force on the inferred states and known forcing function. In particular, it is worth noting that, if a correct estimate is provided for the mass ($\Delta m= 0$), the viscous damping and stiffness errors would simply introduce an additional linear trend in the latent force-velocity and latent force-displacement, respectively. Differently, the presence of a mass error term not only alters these linear trends, but also introduce a dependency on the forcing function and a scaling effect of the friction force. 
	
	\begin{figure}[t!]
		\centering
		\begin{subfigure}{.32\textwidth}
			\includegraphics[height = 4.7cm]{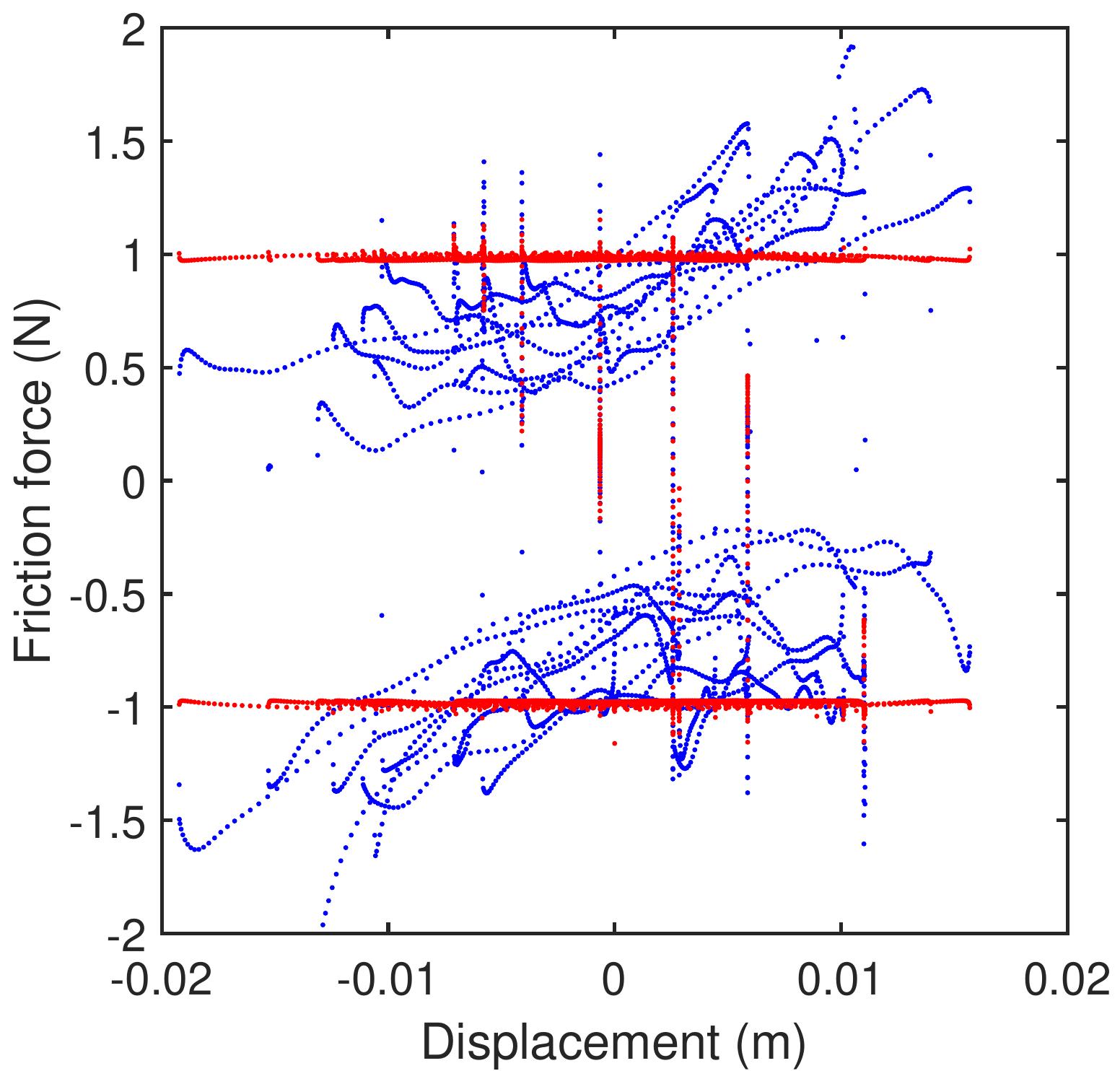}
			\caption{}
		\end{subfigure}
		\hspace{2.25mm}
		\begin{subfigure}{.32\textwidth}
			\includegraphics[height = 4.7cm]{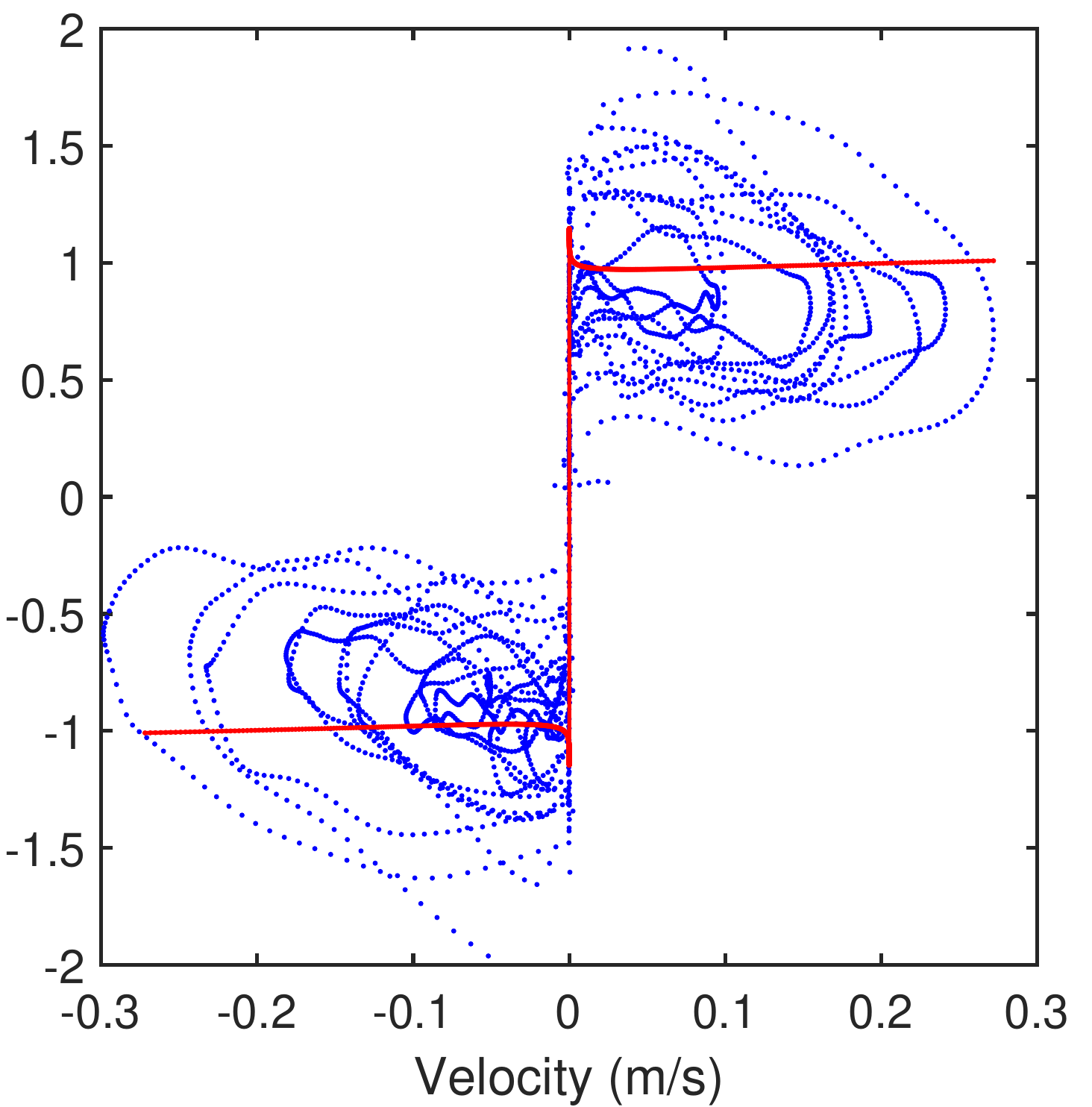}
			\caption{}
		\end{subfigure}
		\hspace{-0.75mm}	
		\begin{subfigure}{.32\textwidth}
			\includegraphics[height = 4.7cm]{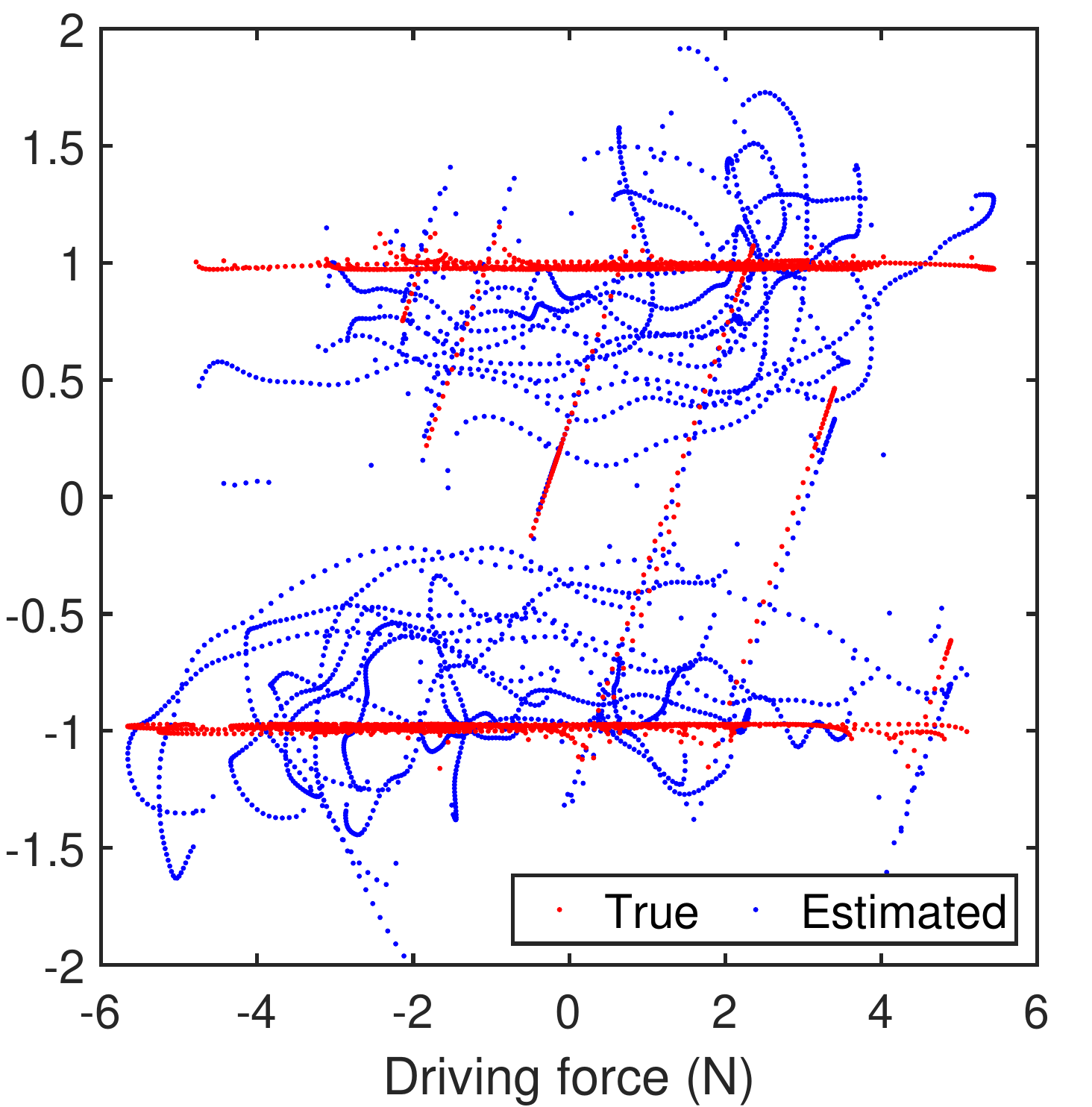}
			\caption{}
		\end{subfigure}
		\caption{Nonlinear friction force vs displacement (\textbf{a}), velocity (\textbf{b}) and driving force (\textbf{c}): comparison between ground truth (in red) and values inferred by using incorrect physical parameters (in blue)}
		\label{fig:8}
	\end{figure}\begin{figure}[t!]
		\centering
		\begin{subfigure}{.32\textwidth}
			\includegraphics[height = 4.7cm]{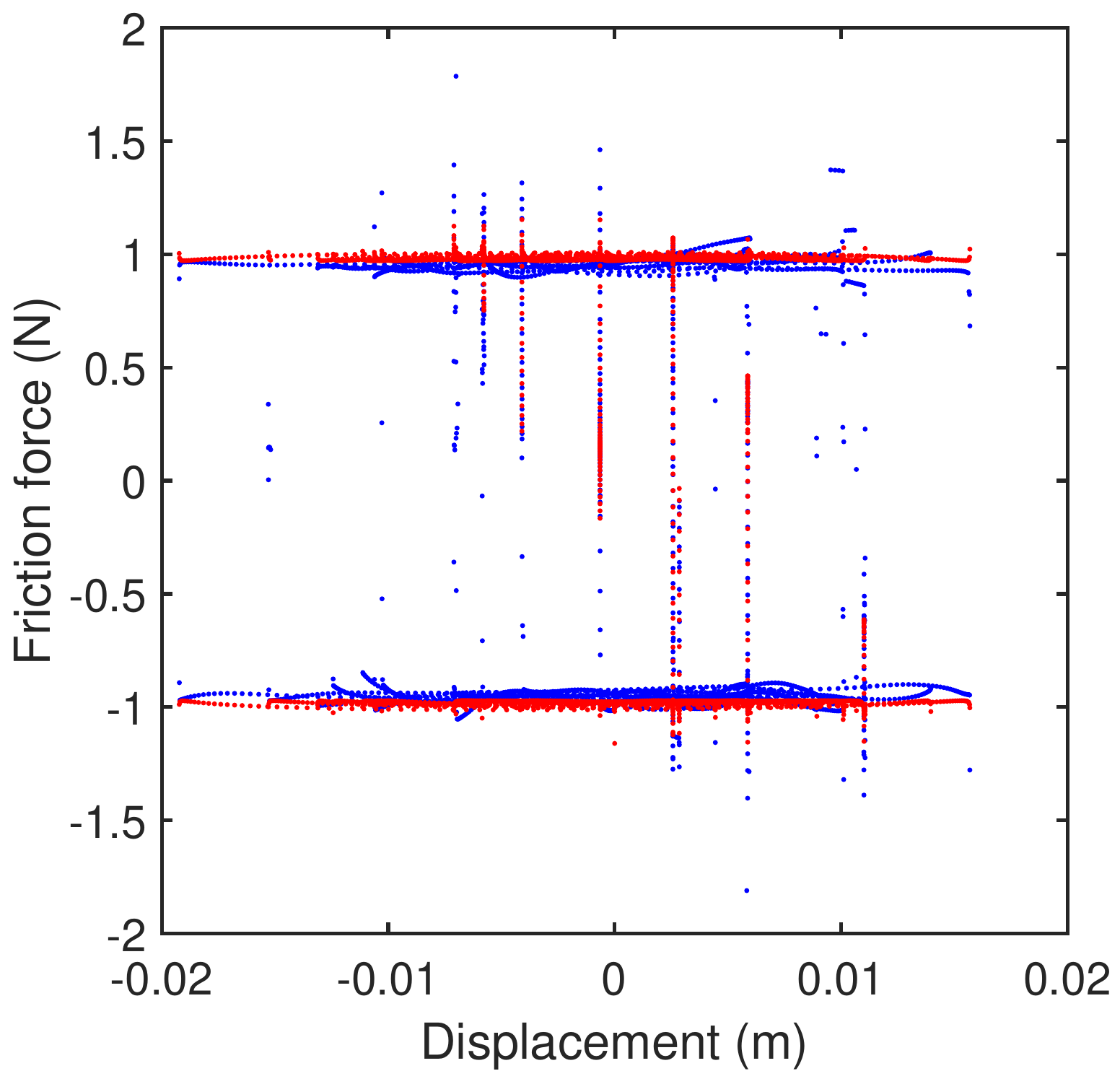}
			\caption{}
		\end{subfigure}
		\hspace{2.25mm}
		\begin{subfigure}{.32\textwidth}
			\includegraphics[height = 4.7cm]{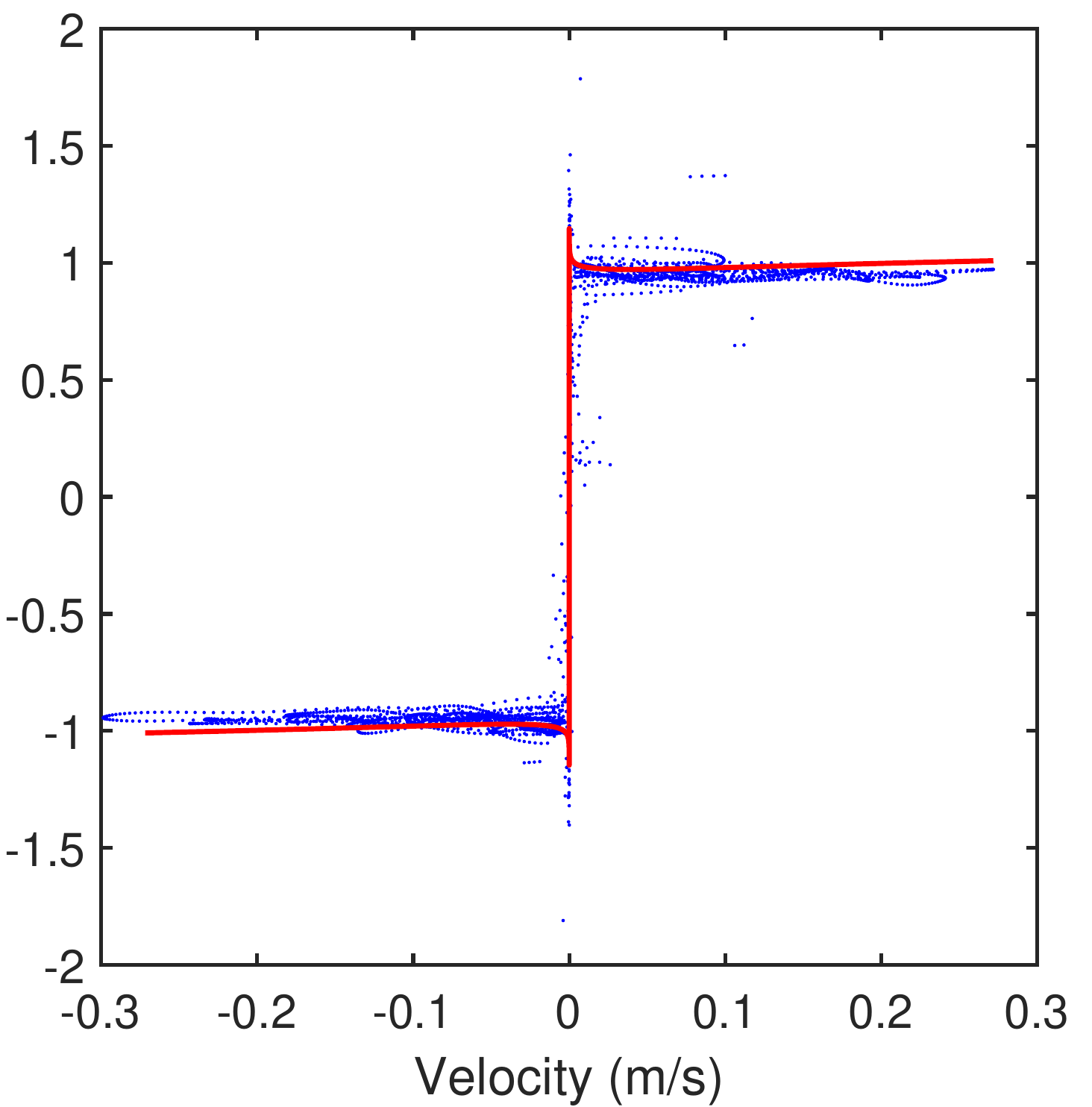}
			\caption{}
		\end{subfigure}
		\hspace{-0.75mm}	
		\begin{subfigure}{.32\textwidth}
			\includegraphics[height = 4.7cm]{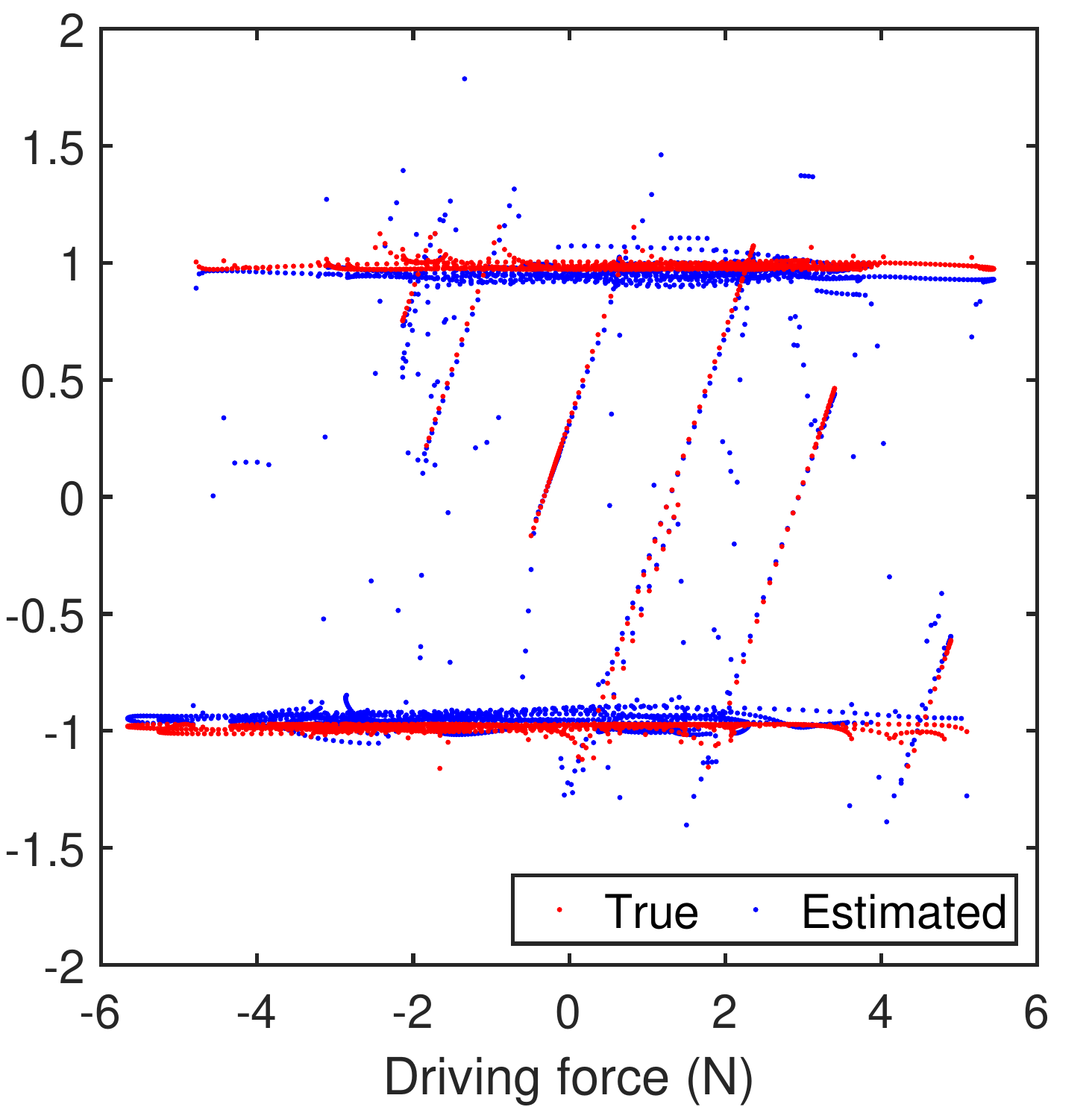}
			\caption{}
		\end{subfigure}
		\caption{Nonlinear friction force vs displacement (\textbf{a}), velocity (\textbf{b}) and driving force (\textbf{c}): comparison between ground truth (in red) and values inferred by using the corrected physical parameters (in blue)}
		\label{fig:9}
	\end{figure}
	
	The latent states and force have been inferred for the dry friction oscillator by considering the incorrect parameters $\hat{m}$, $\hat{c}$ and $\hat{k}$ reported in table \ref{tab5}. The relationships between the latent force estimates and the displacement, velocity and forcing function are illustrated in figure \ref{fig:8}, respectively. In these figures, the presence of the aforementioned linear trends can be observed, along with a general alteration of the latent force patterns due to the presence of a significant mass error. To quantify the linear dependencies expressed in equation (\ref{eq:34}), the latent force can be fitted with a multidimensional linear function, whose coefficients can be retrieved by using a least squared error approach. In order to use all the available data, the symmetry of the friction force $F_f$, and consequently of the latent force $F_L$, with respect to the origin ($z=0$, $\dot{z}=0$, $u=0$) can be exploited for superposing its positive and negative branches, so that they can finally be fitted with the linear function:
	\begin{equation}
		F_L = A_0+A_1z+A_2\dot{z}+A_3u
		\label{eq:35}
	\end{equation}
	By comparing equations (\ref{eq:34}) and (\ref{eq:35}), it is possible to relate the parameter errors to the evaluated linear coefficients as:
	\begin{equation}
		\Delta k = \dfrac{A_1+A_3\hat{k}}{1-A_3} 	
		\qquad\qquad	\Delta c=\dfrac{A_2+A_3\hat{c}}{1-A_3} 
		\qquad\qquad 	\Delta m=\dfrac{A_3\hat{m}}{1-A_3}
		\label{eq:36}
	\end{equation} 
	The estimated parameter errors can finally be added to the initial guesses as indicated in equation (\ref{eq:31}) to retrieve the true parameters of the system. The corrected parameter estimates for the dry friction oscillator case-study are reported in table \ref{tab5}. At this stage, the friction force can be retrieved from the latent force as:
	\begin{equation}
		F_f = \dfrac{1}{1-A_3}\left(F_L-A_1z-A_2\dot{z}-A_3u\right)
		\label{eq:37}
	\end{equation}
	However, it is worth noting that, after the correction of the system parameters, the optimal GP hyperparameters might be different from those initially estimated from VBMC. Therefore, depending on the available computational budget and the relevance of the parameter corrections, it might be preferable to newly infer the optimal hyperparameters and thus the latent states and nonlinear force. This latter approach has been followed in this paper. The inferred friction force is plotted in figure \ref{fig:9} as a function of the displacement, velocity and driving force, respectively, and presents a very good agreement with the ground truth. As reported in table \ref{tab5}, the residual relative errors between the corrected and true parameters are all negligible, despite the very large initial errors. The viscous damping coefficient is the only parameter whose residual error is above 1\%. Although this error would be acceptable for most applications, it is worth mentioning that this slight overestimation is related to the assumed rate-dependent friction model. In fact, as shown in figure \ref{fig:9}b, the selected Dieterich-Ruina's law exhibits an increasing trend with the sliding velocity, which has been identified as an additional viscous damping term. 
	
	\begin{table}[!t]
		\tabcolsep=0pt%
		\TBL{\caption{True, guessed and corrected values of the physical parameters $m$, $c$ and $k$ of the dry friction oscillator. The relative errors of guessed and corrected parameters are referred to the true value. \label{tab5}}}
		{\begin{fntable}
				\centering
				\begin{tabular*}{.95\textwidth}{@{\extracolsep{\fill}}lccccc@{}}\toprule%
					& & \multicolumn{2}{@{}c@{}}{\TCH{Initial guess}}& \multicolumn{2}{@{}c@{}}{\TCH{Corrected}}
					\\\cmidrule{3-4}\cmidrule{5-6}%
					\TCH{Parameters} & \TCH{True value} & \TCH{Value} & \TCH{Rel. err.} & \TCH{Value} & \TCH{Rel. err.} \\\midrule
					\TCH{Mass (kg)} & 1 & 1.2 &  20\% & 0.9550 & 0.50\% \\
					\TCH{Viscous damping (Nsm$^{-1}$)} & 5 & 6 & 20\% & 5.0643 & 1.29\%\\
					\TCH{Stiffness (Nm$^{-1}$)} & 500 & 520 & 4\% & 500.70 & 0.14\%\\
					\botrule
				\end{tabular*}
		\end{fntable}}
	\end{table}

	\subsection{Performance analysis for varying noise levels, observation times and sampling frequencies}
	\label{sec3.6}
	In the previous subsections, the nonlinear system identification performances of the switching GPLFM application to a dry friction oscillator have been investigated by considering a set of observations with a specific level of measurement noise and a fixed number of samples. Therefore, a further step of this investigation consists in analysing how the proposed approach performs when different noise levels, signal durations or sampling frequencies are taken into account. The switching GPLFM is thus applied to the case study investigated in this section for varying $\sigma_n$, $t_f$ and $f_s$, assuming the initial guess reported for the system parameters in table \ref{tab5} and selecting $I=J=3$. The following performance estimators are considered:
	\begin{itemize}
		\justifying
		\item the \textit{nonlinear force identification error} is evaluated as the NMSE of the identified friction force with respect to the ground truth (see equation (\ref{eq:28}));
		\item the \textit{prediction error} aims at evaluating the predictive capabilities of the forward model implemented by considering the best-fitting Dieterich's-Ruina friction law and the corrected parameters estimates. It is calculated as the NMSE of the mass displacement obtained numerically from the forward model with respect to the ground truth;
		\item the \textit{motion regime identification error} is calculated as the relative error between the true and the identified motion regimes. The most probable latent force model is considered as active model at every time step;
		\item the \textit{parameter estimation errors} are the relative errors between estimated and true physical parameters.
	\end{itemize}
	
	\begin{figure*}
		\centering
		\includegraphics[width=0.84\textwidth]{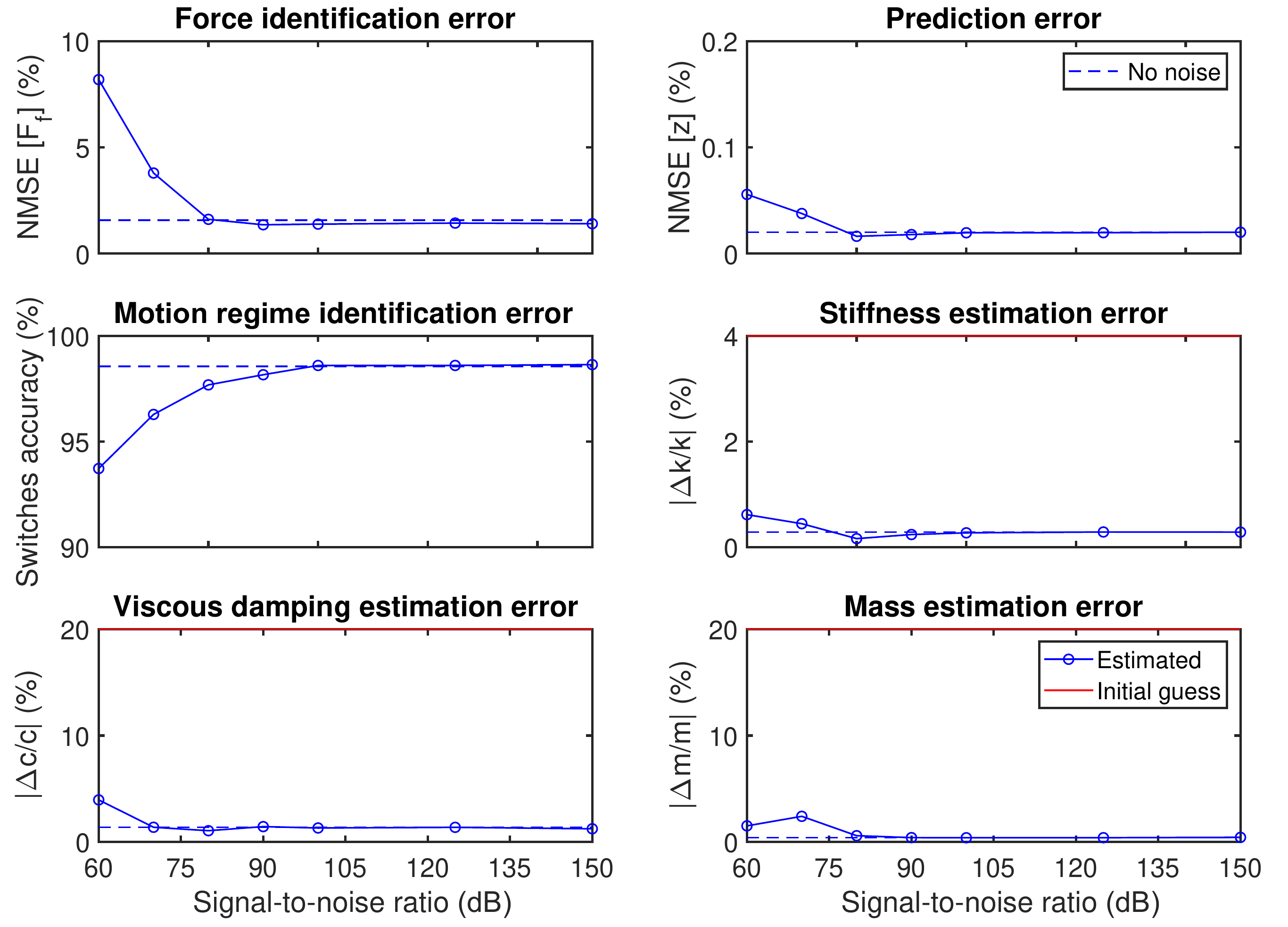}
		\caption{Nonlinear system identification performances of the switching GPLFM ($I=J=3$) applied to the dry friction oscillator case-study for $t_f=5\,$s, $f_s=500$ Hz and varying measurement noise levels}
		\label{fig:10}
	\end{figure*}
	
	\begin{table}[!t]
		\tabcolsep=0pt%
		\TBL{\caption{Conversion between signal-to-noise ratio and standard deviation values for the measurement noise. \label{tab6}}}
		{\begin{fntable}
				\small
				\centering
				\begin{tabular*}{\textwidth}{@{\extracolsep{\fill}}lccccccc@{}}\toprule%
					\TCH{\textbf{SNR (dB)}} \quad & 60 & 70 & 80 & 90 & 100 & 125 & 150 \\
					\TCH{$\boldsymbol{\sigma_n}$\textbf{ (mm)}} & \hspace{2mm} 0.0786 \hspace{2mm} & \hspace{2mm} 0.0249 \hspace{2mm} & $7.862 \times 10^{-3}$ & $2.486 \times 10^{-3}$ & $7.862 \times 10^{-4}$ & $4.421 \times 10^{-5}$ & $7.862 \times 10^{-3}$\\
					\botrule
				\end{tabular*}
		\end{fntable}}
	\end{table}
	
	\begin{figure}[!t]
		\centering
		\includegraphics[width=0.84\textwidth]{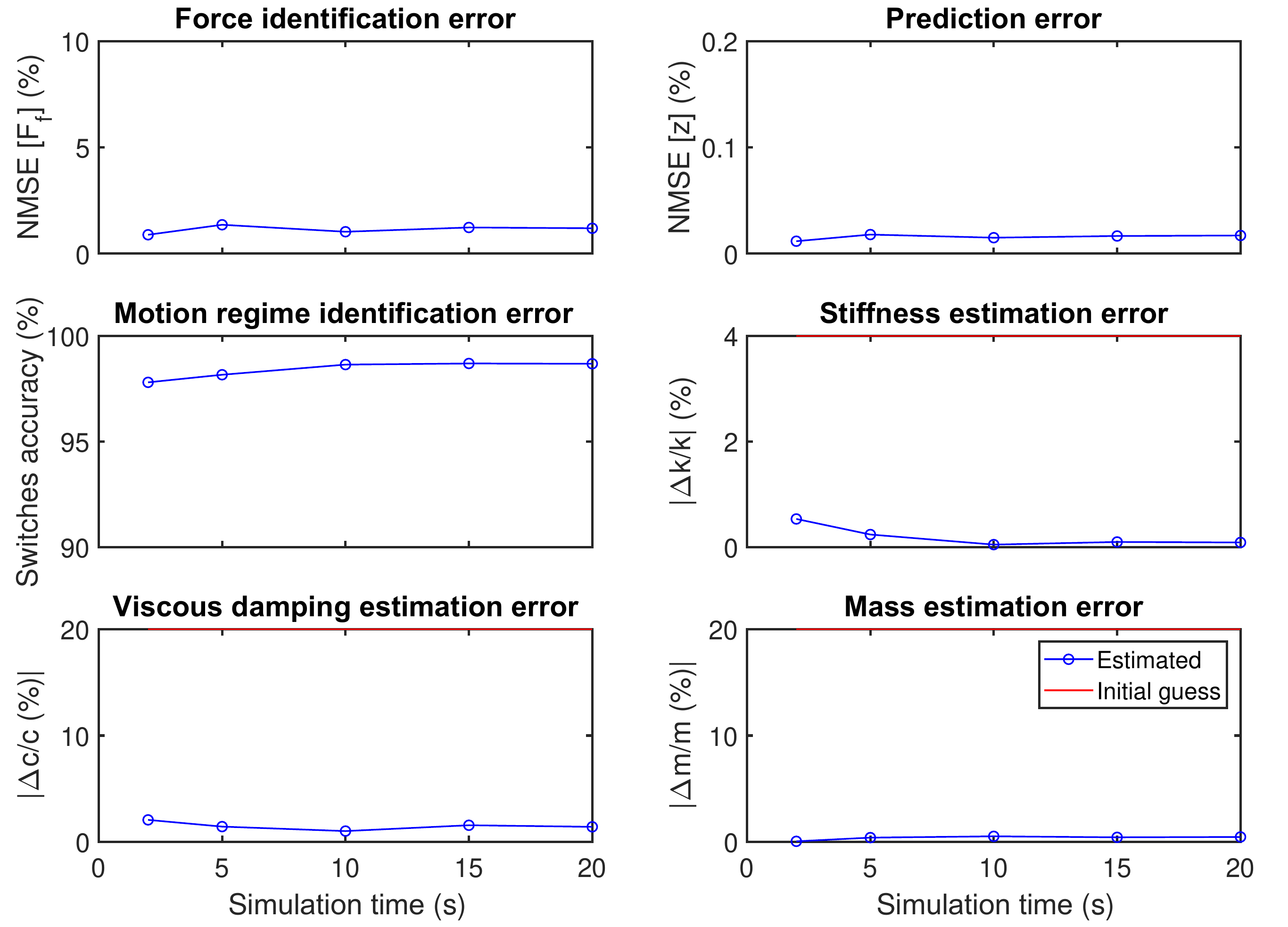}
		\caption{Nonlinear system identification performances of the switching GPLFM ($I=J=3$) applied to the dry friction oscillator case-study for SNR $=80$ dB, $f_s=500$ Hz and varying simulation times}
		\label{fig:11}
	\end{figure}
	\begin{figure}[!h]
		\centering
		\includegraphics[width=0.84\textwidth]{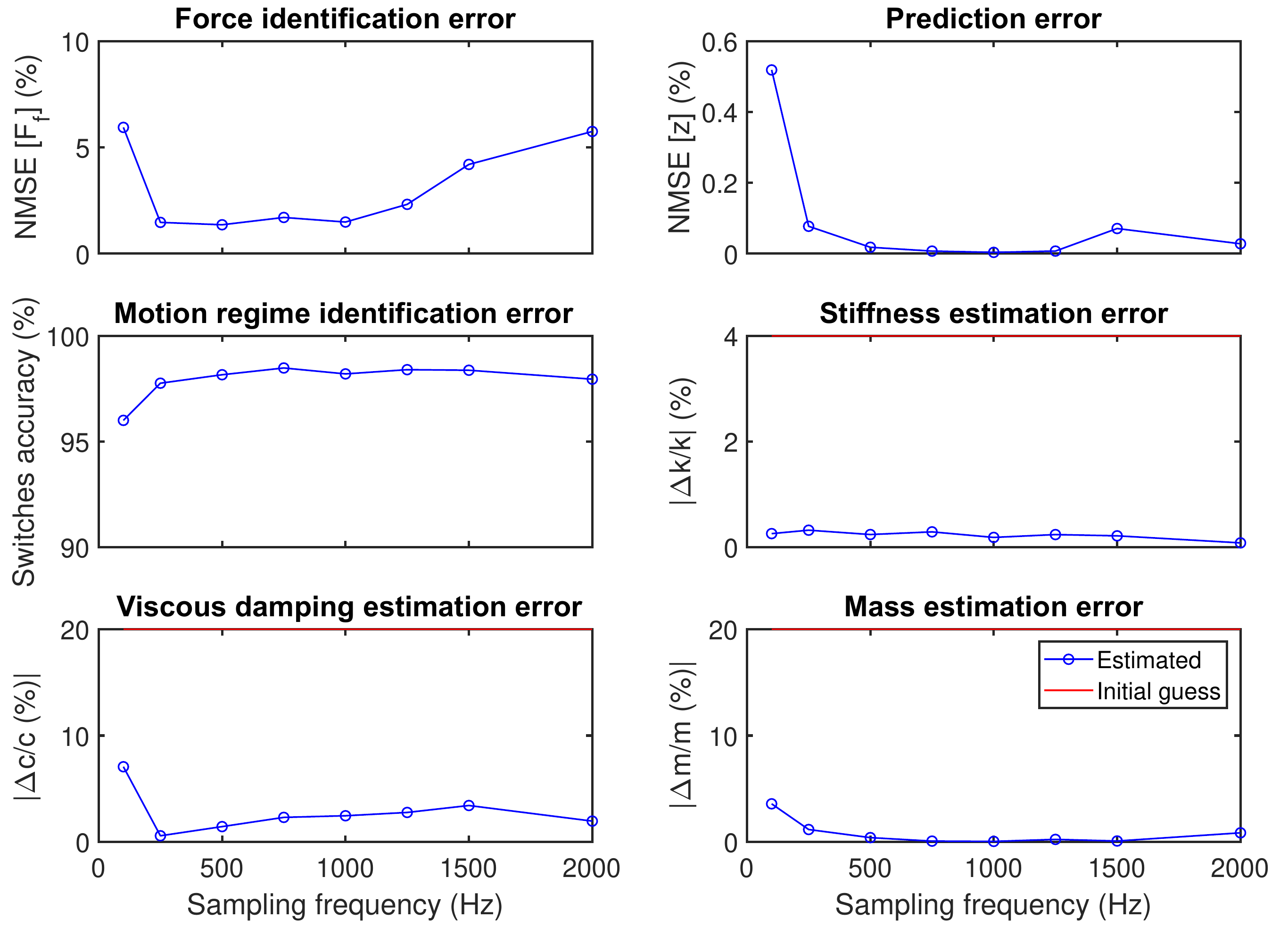}
		\caption{Nonlinear system identification performances of the switching GPLFM ($I=J=3$) applied to the dry friction oscillator case-study for SNR $=80$ dB, $t_f=5\,$s and varying sampling frequencies}
		\label{fig:12}
	\end{figure}
	
	The switching GPLFM performances have been investigated for seven different levels of measurement noise (SNR = [60, 70, 80, 90, 100, 125, 150] dB) and for a noise-free measurement (SNR = $\infty$), while maintaining the final time and the sampling frequency of the observations set to $t_f=5\,$s and $f_s=500$ Hz. The corresponding values of the standard deviation $\sigma_n$ are reported in table \ref{tab6}. The performances indexes are plotted in figure \ref{fig:10}. In this figure, it is possible to observe that switching GPLFM presents similarly good performances with respect to all the measured errors when the signal-to-noise ratio is equal or above to 80 dB. For higher noise levels, the main effect appears to be a decrease of the motion regime identification accuracy, which also leads to lower identifications scores. Nonetheless, all the scores maintain generally acceptable value for SNRs larger than 60 dB.
	
	A further performance analysis has been carried out maintaining the noise levels and the sampling frequency set to SNR$=80$ dB and $f_s=500$ Hz, while varying the simulation time between 2 and 20$\,$s. The resulting error indexes, reported in figure \ref{fig:11}, are substantially constant across different observed times. Therefore, it can be deduced that that the switching GPLFM can offer stable performances for varying number of samples, as long as the time step among the observations is maintained constant.
	
	The last investigation has been performed by varying the sampling frequency of the measurements between 100 and 2000 Hz; this corresponds to a variation of the fixed time step between 0.0005 and 0.01$\,$s. The SNR and time span of the measurements have been set to 90 dB and 5$\,$s, respectively. Observing the resulting scores, illustrated in figure \ref{fig:12}, it is clear that the sampling frequency has a larger impact on the performances than the simulation time. While large errors are displayed for most of the investigated parameters when $f_s = 100$ Hz, indicating that the switching GPLFM cannot perform well at very low sampling frequencies, very good scores are obtained when the sampling frequency is increased to 250 Hz. At the other of the investigated range, it is very interesting to observe that the nonlinear force identification error increases significantly at high sampling frequency, indicating the oversampling issues might occur when applying the GPLFM. On the other hand, the prediction, motion regime identification and parameter estimation errors do not appear to be affected by oversampling.
	
	\section{Experimental case-study: single-storey frame with a brass-to-steel contact}
	\label{sec4}
	The applicability of the switching latent restoring force model is further investigated by considering an experimental case-study involving a harmonically base-excited single-storey frame with a metal-to-metal contact. The goal is the identification of the nonlinear friction force generated in the contact under the action of a normal load and the reconstruction of the friction force-velocity relationship for varying normal load amplitudes.
	
	\begin{figure}[h!]
		\centering
		\begin{subfigure}{.55\textwidth}
			\centering
			\includegraphics[width=\textwidth]{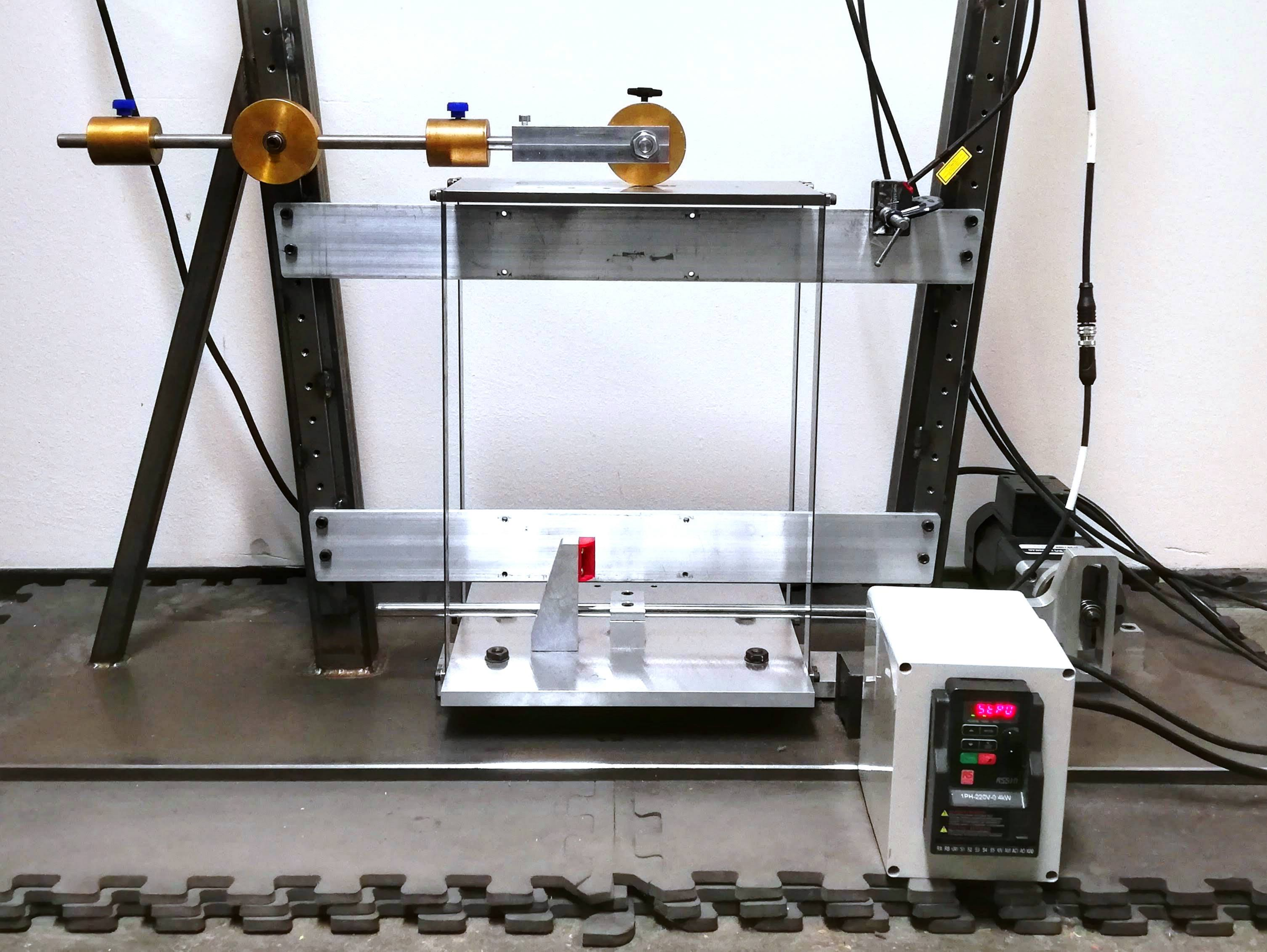}
			\caption{}
		\end{subfigure}
		\hspace{12mm}
		\begin{subfigure}{.23\textwidth}
			\centering
			\vspace{10mm}
			\includegraphics[width=\textwidth]{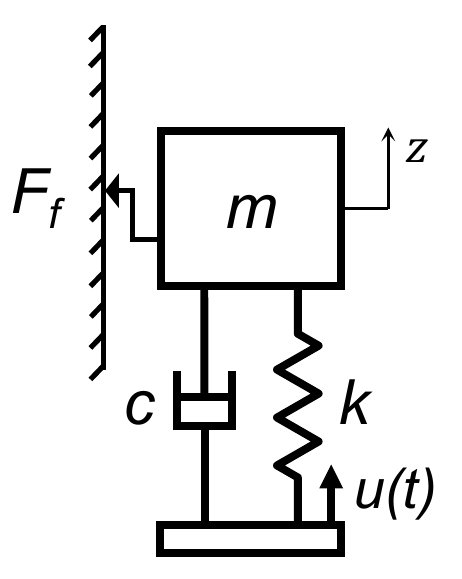}
			\vspace{4mm}
			\caption{}
		\end{subfigure}
		\caption{Picture (\textbf{a}) and mechanical model (\textbf{b}) of the test rig: base-excited single-store frame with a brass-to-steel contact}
		\label{fig:13}
	\end{figure}
	
	\subsection{Test rig and mechanical model}
	\label{sec4.1}
	The experimental tests have been carried out on the rig shown in figure \ref{fig:13}a, which is briefly described in what follows. The main structure is a single-storey frame made up of two metal plates connected by four doubly-bolted thin metal bars. The bottom plate is excited by the alternate motion of a shaft, which is connected to an electric motor through a Scotch-yoke mechanism. This results in an approximatively harmonic base excitation, whose driving frequency can be specified by selecting the rotation speed of the motor. A brass disc is placed on the top steel plate to create a friction contact. The disc is mounted on a counterweight system pinned to the external frame, so that the normal force exerted on the top plate, and therefore the friction force amplitude, can be adjusted by shifting the weights along the counterweight axis. The displacement of the two plates is measured by laser position sensors, thus providing input and output measurements during the tests. A more detailed explanation of the setup and testing procedure can be found in \cite{Marino2020}. 
	
	In the above publication, it was shown that the dynamic behaviour of the single-storey frame is well reproduced by the single-degree-of-freedom mass-spring model shown in figure \ref{fig:13}b when driving frequencies up to at least 8 Hz are considered. Moreover, the experimental results obtained by \citeauthor{Marino2020} for the dynamic response of this structure agree well with the theoretical results obtained analytically and numerically by considering the Coulomb friction model. Since Coulomb's law introduces a discontinuity in the friction force at zero sliding velocity, it is reasonable to assume that the single-storey frame behaves as a discontinuous nonlinear system, thus representing a suitable case-study for the proposed method.
	
	The governing equation of the mechanical system schematised in figure \ref{fig:13}b can be written as:
	\begin{equation}
		m\ddot{z}+c\dot{z}+kz+F_f(z,\dot{z}) = ku(t)+c\dot{u}(t)
		\label{eq:38}
	\end{equation}
	where $u(t)$ and $\dot{u}(t)$ are the displacement and velocity of the base. Differently from the mechanical system investigated in section \ref{sec3}, where the known driving force was directly applied to the mass, the input term of equation (\ref{eq:38}) is made up of two terms, both dependent on generally unknown physical parameters, and the derivative of input base excitation must also be provided. The probabilistic model implemented in section \ref{sec3.1} can be easily adapted to deal with a base-excited oscillator. Specifically:
	\begin{itemize}
		\justifying
		\item in the state-space model of the system, the input vector will be written as $\mathbf{u}=\begin{bmatrix}u &\dot{u}\end{bmatrix}^\top$ and the matrix $B_{c,s}$ will be formulated as: 
		\begin{equation}
			B_{c,s} = \begin{bmatrix}0 & 0\\-\frac{k}{m} &-\frac{c}{m}\end{bmatrix}
			\label{eq:39}
		\end{equation}
		\item in the predicted mean value of the latent restoring force from equation (\ref{eq:27}), the term $u_{t-1}$ will be replaced by $ku_{t-1}+c\dot{u}_{t-1}$.
	\end{itemize}
	Nonetheless, while the formulation of the switching latent force model is not heavily affected by the presence of the base excitation, difficulties may arise regarding the estimation of the physical parameters and input vector. 
	
	The first problem is related to the non-identifiability of the mass. In fact, rewriting equation (\ref{eq:38}) as:
	\begin{equation}
		\ddot{z}=\dfrac{k}{m}u(t)+\dfrac{c}{m}\dot{u}(t)-\dfrac{k}{m}z-\dfrac{c}{m}\dot{z}-\dfrac{1}{m}F_f(z,\dot{z})
		\label{eq:40}
	\end{equation}
	it can be observed that, in the right-hand side, all the terms except the unknown friction force are multiplied by either $k/m$ or $c/m$. Therefore, if the mass value is changed from its true value $m$ to an incorrect estimate $\hat{m}$ while keeping these ratios constant, the only effect on the identified states and latent force is that the latter would be scaled by a factor $\hat{m}/m$. Obviously, since the friction force amplitude is unknown, this scaling factor cannot be determined from the identified latent force. In conclusion, independently of the system identification approach considered, it is not possible to identify the mass parameter of a base-excited system from its base and mass displacements. In this contribution, the value $m=3.0799$ kg will be assigned to the mass. This value has been obtained by performing a hammer test on the top plate of the frame after removing the counterweight system and, therefore, the friction contact. The experimental frequency response function has thus been fitted with that of a mass-spring-damper system in the frequency range 0-6 Hz, retrieving the optimal parameters of the system with a least squares method. The so-obtained physical and modal parameters of the system are reported in table \ref{tab7}, where they are compared to the parameters estimated by fitting the identified latent force with the linear function:
	\begin{equation}
		F_L = A_0+A_1(z-u)+A_2(\dot{z}-\dot{u})
		\label{eq:41}
	\end{equation}
	and correcting the initial estimates by adding $\Delta k=A_1$ and $\Delta c=A_2$, according to the procedure from subsection \ref{sec3.2}.
	
	A further issue in the application of the switching GPLFM to this experimental case-study is that the base displacement and velocity must be provided as an input. However, although the fundamental frequency of the base excitation is selected by the user, only noisy measurements of the bottom plate displacements are available from the laser sensors. To recover the driving functions $u(t)$ and $\dot{u}(t)$, the following state-space model has been implemented from Newton's laws of motion:
	\begin{subequations}
		\begin{empheq}[left = \empheqlbrace]{align}
			\begin{bmatrix} u_t \\ \dot{u}_t \end{bmatrix} &= \begin{bmatrix} 1 & \Delta t \\ 0 & 1 \end{bmatrix}\begin{bmatrix} u_{t-1} \\ \dot{u}_{t-1} \end{bmatrix} + \begin{bmatrix} \frac{1}{2}\Delta t^2 \\ \Delta t \end{bmatrix}a_{t-1}, \qquad 
			a_{t-1}\sim\mathcal{N}(0,\sigma_a^2)\\
			y_{u,t} &= \begin{bmatrix}1 & 0 \end{bmatrix}\begin{bmatrix} u_t \\ \dot{u}_t \end{bmatrix}+v_{u,t}, \hspace{2.4cm}
			v_t\sim\mathcal{N}(0,\sigma_{u}^2)
		\end{empheq}
		\vspace{-3mm}
		\label{eq:42}
	\end{subequations}\\
	by assuming that an unknown constant acceleration $a_t$, modelled as a zero-mean normal distribution with variance $\sigma_a^2$, acts between the time steps $t$ and $t+1$. The above state-space model has then been computed via Kalman filtering and RTS smoothing, by considering the optimal values of the hyperparameters inferred by VBMC. It is worth mentioning that, while the proposed procedure can be conveniently applied in the assumption that the base motion is not affected by either the dynamic response of the structure or the friction force, a more general approach would consist in coupling the above state-space model with the augmented state-space representation of the latent force models.
	
	\begin{figure}[!h]
		\centering
		\includegraphics[width=0.84\textwidth]{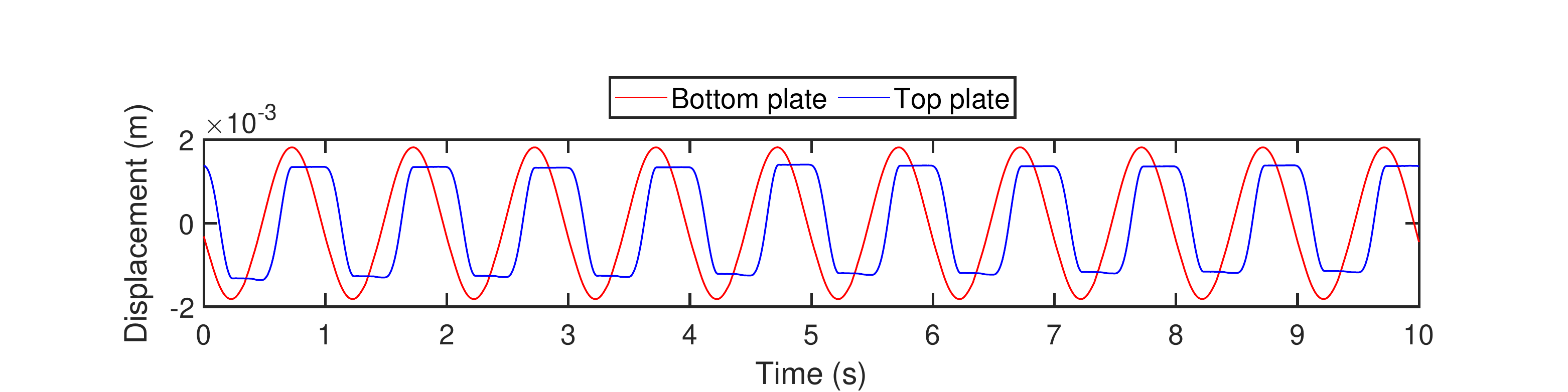}\hspace{.5cm}
		\caption{Measured displacements of the bottom and top plates of the single-storey frame under harmonic excitation. The driving frequency is set at 1 Hz and a normal load of 5.5 N is applied to the top plate}
		\label{fig:14}
	\end{figure}
	
	\subsection{Inference and results}
	Nonlinear system identification has been performed by applying the proposed switching GPLFM ($I=J=3$) to the experimental data obtained by setting a fundamental driving frequency of 1 Hz and a normal force equal to 5.5 N. The displacements of the two plates, reported in figure \ref{fig:14}, have been recorded for 10$\,$s with a sampling frequency of 250 Hz.
	
	The estimated physical and modal parameters are listed in table \ref{tab7}. In particular, the estimated stiffness and natural frequency agree well with the measured values from hammer testing, while the estimated viscous damping and the corresponding damping ratio are slightly overestimated. Nonetheless, viscous damping has negligible values in both cases, in agreement with the results presented by \cite{Marino2020}. 
	
	\begin{figure*}
		\centering
		\includegraphics[width=0.9\textwidth]{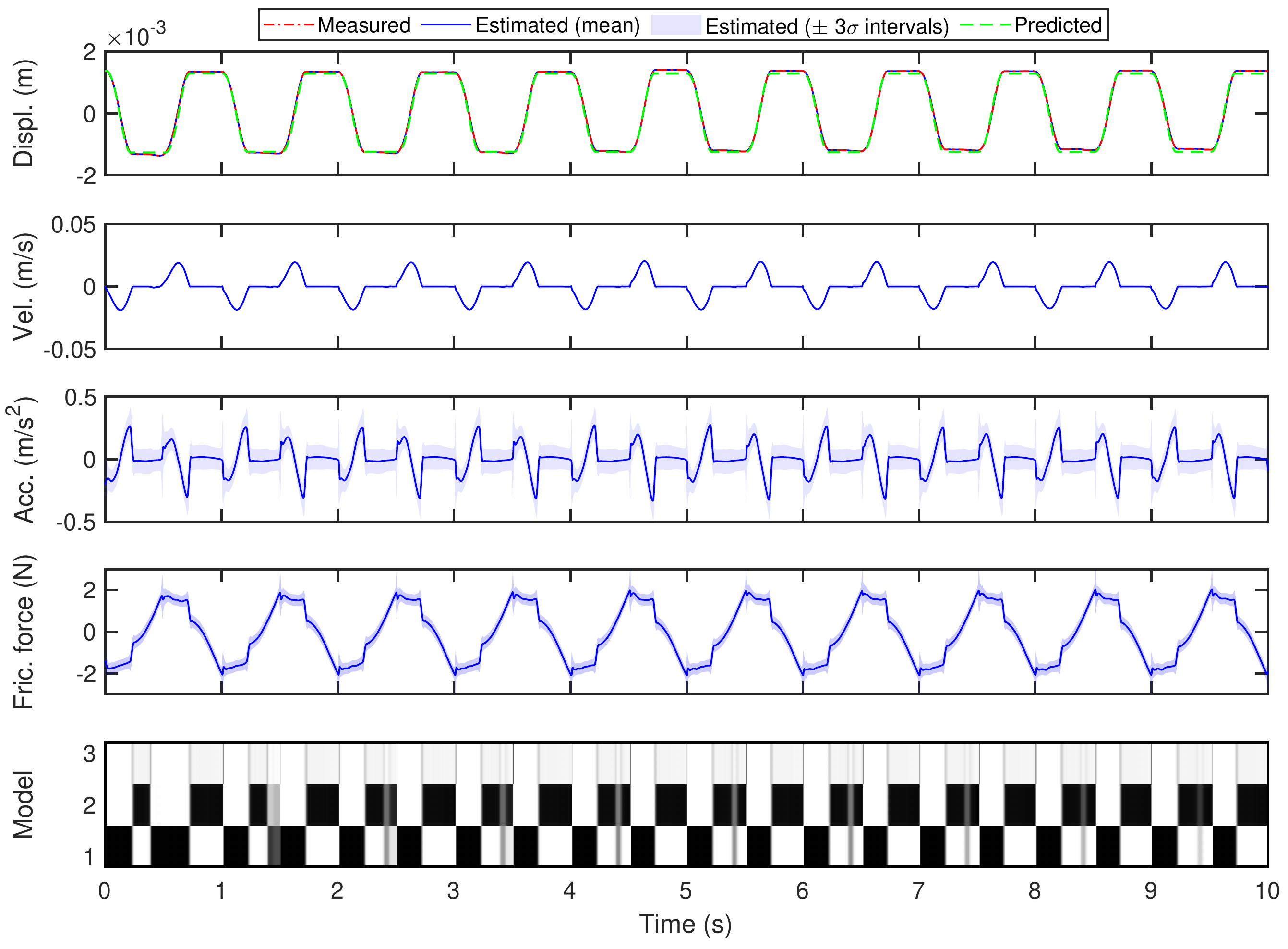}
		\caption{Latent states, acceleration, nonlinear friction force and model probabilities inferred by switching GPLFM ($I=J=3$). Models 1,2 and 3 stands for sliding, sticking and resetting models, respectively}
		\label{fig:15}
	\end{figure*}
	\begin{table}[!t]
		\tabcolsep=0pt%
		\TBL{\caption{Physical and modal parameters of the single-storey frame: estimated values from hammer testing vs initial guess and corrected values in the identification procedure. \label{tab7}}}
		{\begin{fntable}
				\centering
				\begin{tabular*}{.85\textwidth}{@{\extracolsep{\fill}}lccc@{}}\toprule%
					\TCH{Parameters} & \TCH{Hammer test} & \TCH{Initial guess} & \TCH{Corrected}  \\\midrule
					\TCH{Mass (kg)} & 3.0799 &  &  \\
					\TCH{Viscous damping (Nsm$^{-1}$)} & 0.3004 & 5 & 0.6691\\
					\TCH{Stiffness (Nm$^{-1}$)} & 1197 & 1250 & 1191 \\				
					\TCH{Natural frequency (Hz)} & 3.1374 & 3.2063 & 3.1298 \\
					\TCH{Modal damping ratio} & 0.0018 & 0.0403 & 0.0055 \\
					\botrule
				\end{tabular*}
		\end{fntable}}
	\end{table}
	
\begin{figure*}
		\centering
		\includegraphics[width=0.62\textwidth]{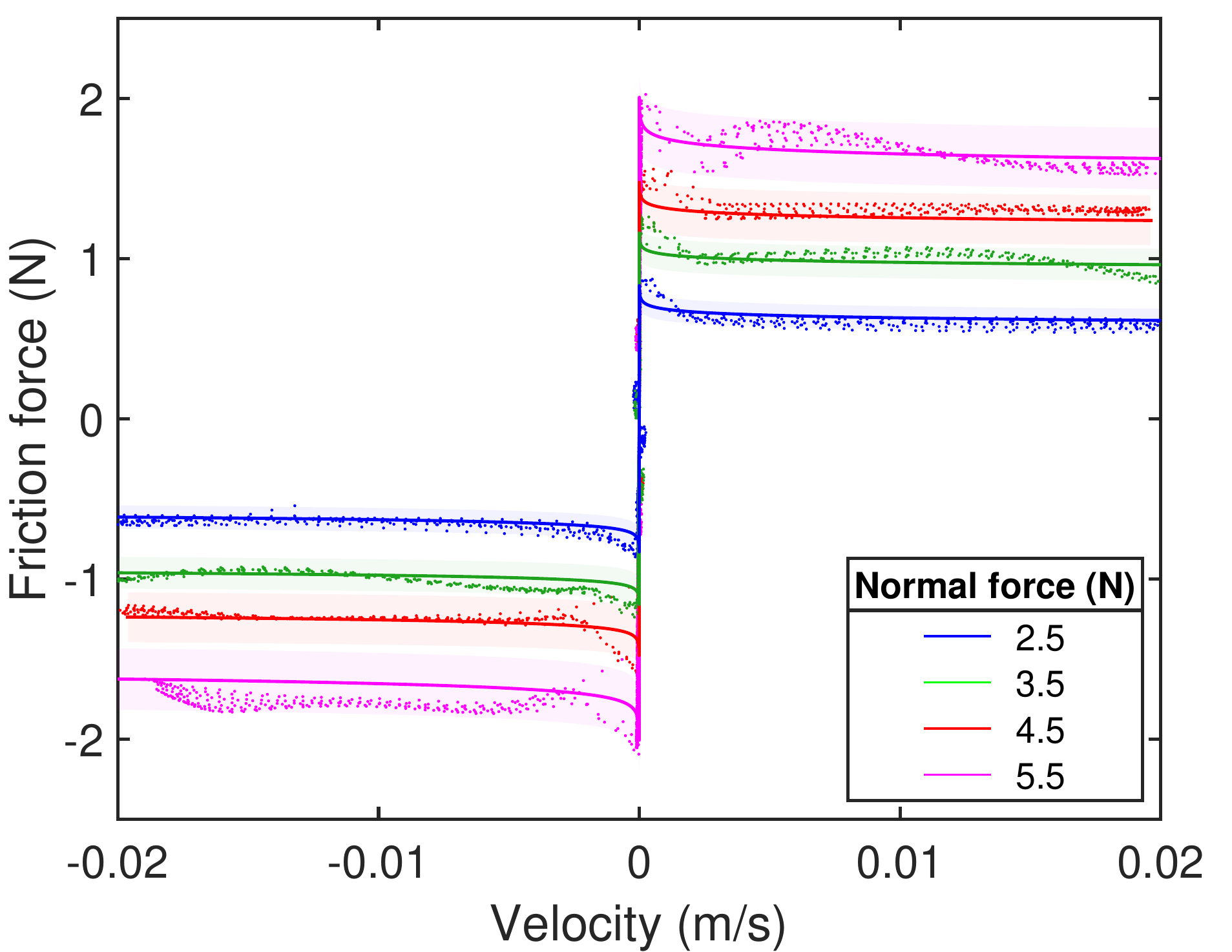}
		\caption{Friction force vs velocity for varying applied normal loads: points estimates (dots), fitted friction laws (continuous lines) and $\pm 3\sigma$ confidence intervals (shaded areas)}
		\label{fig:16}
	\end{figure*}

	The identified latent states and friction force are reported in figure \ref{fig:15}, along with the sequence of model probabilities. In this figure, it is possible to observe that the estimated displacements present an excellent agreement with the experimental observations, scoring a NMSE equal to 4.889$\times10^{-6}$. The displacement of the top plate is a quite regular two-stops stick-slip motion. As shown in the bottom frame of figure \ref{fig:15}, the transitions between the sliding and sticking phases are generally well captured by the switching latent force model. Nonetheless, during the stops occurring at $z<0$ in the top plate motion, slow variations can be observed before the onset of the subsequent sliding phases. This particular behaviour clearly renders more difficult the identification of the motion regime, leading to the presence of grey areas in the model probabilities corresponding to these stops. The identified friction force does not appear to be affected by the larger uncertainty that characterises negative stops; in fact, regular patterns are observed during each sticking phase. The sliding friction force is also characterised by generally regular behaviours, with a slight decrease in amplitude during each sliding phase. This trend can also be visualised in the latent friction force-velocity point estimates reported (with magenta dots) in figure \ref{fig:16}.
	
	The friction force-velocity estimates have been fitted with the steady-state Dieterich-Ruina's law introduced in equation (\ref{eq:24}a); the value of the static friction force has been determined as described in subsection \ref{sec3.4}. In figure \ref{fig:16}, it can be observed that both branches of the friction force-velocity curve follow a monotonic behaviour, characterised by a decrease of the friction force amplitude for increasing absolute values of the sliding velocity. This particular pattern, known in the friction literature as Stribeck effect, is reproduced by the Dieterich-Ruina's law due to the very small identified values for the parameters $a$ and $c$ (see also \cite{Cabboi2022}). 
	
	To evaluate the predictive capabilities of the switching GPLFM, the dynamic response of the system has been simulated numerically by considering the fitted friction model and the estimated physical parameters (see table \ref{tab7}). The measured and simulated mass displacements present a very good agreement, also visible in the top frame of figure \ref{fig:15}; their comparison yields a NMSE score of 0.4316\%.
	
	Finally, the proposed approach has been applied to experimental data obtained for different normal force levels, ranging from 2.5 to 5.5 N. The obtained friction force-velocity estimates, along with the corresponding fitted friction laws, are reported in figure \ref{fig:16}. It can be observed that the friction force-velocity curves present similar trends across the different applied normal forces. In addition, the curves are almost equally spaced, suggesting that the friction force amplitude increases linearly with the normal force, as expected. The predictive performances of the fitted friction models do not present significant variations are substantially maintained for the different values of $N$, with NMSE scores never exceeding 1\% among the observed cases.
	
	\section{Conclusion}
	\label{sec5}
	The work presented in this paper demonstrates that switching Gaussian process latent force models can be effectively used for identifying complex dynamical systems with discontinuous nonlinearities and/or multiple motion regimes. By using a set of noisy observations of the system's response to a known input force, GPLFMs can infer the time histories of the latent states and the nonlinear force, without requiring prior knowledge of the functional form of the latter. Nonetheless, standard GPLFMs, where the latent nonlinear force is modelled by a single Gaussian process, cannot easily handle discontinuous nonlinear forces, such as those generated by frictional contact. In this paper, this identification problem has been tackled by introducing a switching GPLFM, where multiple GP latent forces can be used to model the nonlinear force across different motion regimes, and their priors can be updated using assumed density filtering and an expectation-correction smoothing algorithm. Additionally, a resetting model has been included among the latent force models to achieve discontinuities in the nonlinear force. The model transitions are also inferred in a Bayesian manner, along with the latent states and nonlinear force, by using the state-of-the-art methodology for switching linear dynamical systems. 
	The proposed switching GPLFM has been applied to two case-studies, including a simulated dry friction oscillator and an experimental setup consisting in a single-storey frame with a brass-to-steel contact. In both cases, excellent results were obtained in terms of the identified nonlinear and discontinuous friction force for varying (i) normal load amplitudes in the contact; (ii) measurement noise levels and (iii) number of samples in the datasets.
	
	The switching GPLFM can offer several advantages over pure data-driven approaches for nonlinear system identification. By embedding noisy observations and a partially-known physics-based model in the probabilistic model,  a physics-enhanced machine learning model is obtained. This method allows not only the recognition of discontinuities in the time series, but also the introduction of physical constraints into the model, such as those imposed during the sticking phases in dry friction oscillators. Therefore, this approach is highly \textit{interpretable},  since it yields transparent predictions  based on the explicit inclusion of a large amount of information deriving from physical and engineering knowledge. Most importantly,  even in the presence of a limited dataset, the proposed switching GPLFM yields an increased generalisation performance with respect to extrapolation and observational biases, by enforcing explicitly causal relationships and physics-based constraints.  Further,  physically inconsistent or implausible predictions can be easily detected by employing the identified discontinuous nonlinear force (e.g., the identified friction force-velocity curves and static friction force values estimates for the dry friction problem) in a \textit{forward model} under input data not included in the training dataset, to assess discrepancies with the corresponding output measurements. 
	Most importantly this approach is \textit{explainable}, since the results obtained are easy to understand by humans. They can be expressed in terms of robust physics-based features, such as the friction force-velocity relationship, and the results also include an assessment of the remaining uncertainty in the robust features and on the system states predicitons. 
	
	The proposed approach is generally applicable to the analysis of engineering systems subject to a single nonsmooth nonlinearity that can be approximated by a single degree-of-freedom model. 
	Future work will focus on: (1) extending the current formulation to multi degree-of-freedom models, particularly those where contacts and/or other nonlinearities may occur simultaneously on different masses, and (2) accounting for more complex friction models, characterised by dependencies on further state variables (internal states, temperature, etc.) and/or by a time-evolution of the states in sticking regime.
	It is also worth mentioning that switching GPLFM, requires the introduction of multiple latent force models and the use of Gaussian mixtures. Therefore, it requires a larger computational cost compared to the application of a standard GPLFM. Nonetheless, the performance analysis presented in this paper has shown that a small number of Gaussian components is often sufficient to obtain significant improvements in the identification accuracy of discontinuous nonlinear force.

	\begin{Backmatter}
		
		
		\paragraph{Author Contributions}
		Conceptualization: all authors; Data curation: L.M.; Investigation: L.M.; Methodology: L.M.; Project management: A.C.; Supervision: A.C.; Writing original draft: L.M.; Writing – review \& editing: all authors. All authors approved the final submitted draft.
		
		\paragraph{Competing Interests}
		None.
		
		\paragraph{Data Availability Statement}
		MATLAB codes and experimental data used in this paper are available on GitHub: \verb+\url{https://github.com/l-marino/switching-GPLFM/}+.
		
		\paragraph{Ethical Standards}
		The research meets all ethical guidelines, including adherence to the legal requirements of the study country.
		
		\paragraph{Funding Statement}
		This work received no specific grant from any funding agency, commercial or not-for-profit sectors.
		
		
		\bibliographystyle{apalike}

	\end{Backmatter}
	
\end{document}